\documentclass[aps,prl,reprint, superscriptaddress]{revtex4-1}
\usepackage[section]{placeins}

\usepackage{lineno}
\usepackage{graphicx}  % needed for figures
\usepackage{dcolumn}   % needed for some tables
\usepackage{bm}        % for math
\usepackage{amssymb}   % for math
\usepackage{amsmath,stmaryrd}
\usepackage{blkarray, multirow, graphicx, diagbox, color, colortbl}
\usepackage[dvipsnames]{xcolor}
\usepackage{bbm, bbold}
\usepackage{glossaries}
\usepackage{algorithm}
\usepackage{algpseudocode}
\usepackage{hyphenat}
\usepackage{ifthen}
\usepackage[colorlinks, linkcolor = blue, citecolor = blue, filecolor = black, urlcolor = blue]{hyperref}
\usepackage{dsfont}
\usepackage{xkeyval}
\usepackage{moreverb}
\usepackage{rotating}
\usepackage{wrapfig}
\usepackage{slashbox}
\usepackage{xspace}
\usepackage{nicefrac}
\usepackage[]{units}
\usepackage{physics}
\usepackage{booktabs}
\usepackage{braket}
\usepackage[inline]{enumitem}
\usepackage{tabto}
\usepackage{listings}
\usepackage[normalem]{ulem}
\usepackage{xstring}
\def\ReplaceStr#1{%
	\IfSubStr{#1}{p}{%
		\StrSubstitute{#1}{p}{.}}{#1}}
\usepackage[english]{babel}

\usepackage[caption=false]{subfig}
\usepackage[capitalise]{cleveref} %
\usepackage{chemformula}
\usepackage{algorithm}

\newcommand{\nodagger}[0]{{\vphantom{\dagger}}}

\newcommand{\updated}[1]{\textcolor{black}{#1}}

\newacronym{OBC}{OBC}{open boundary condition}
\newacronym{PBC}{PBC}{periodic boundary condition}
\newacronym{DSP}{DSP}{dissipative state preparation}
\newacronym{MPS}{MPS}{matrix\hyp product state}
\newacronym{MPO}{MPO}{matrix\hyp product operator}
\newacronym{ED}{ED}{exact diagonalization}
\newacronym{BEC}{BEC}{Bose-Einstein condensate}
\newacronym{DPT}{DPT}{dissipative phase transition}
\newacronym{CDW}{CDW}{charge density wave}
\newacronym{1D}{1D}{one dimensional}
\newacronym{LSE-TDVP}{LSE-TDVP}{local subspace expansion time-dependent variational principle}
\newacronym{TDVP}{TDVP}{time-dependent variational principle}
\newacronym{ONB}{ONB}{orthonormal basis}
\newacronym{2TDVP}{2TDVP}{two-site time-dependent variational principle}
\newacronym{HEOM}{HEOM}{hierarchy of equations of motion}
\newacronym{DMRG}{DMRG}{density matrix renormalization group}
\newacronym{SL}{SL}{symmetrically\hyp localized}
\newacronym{TN}{TN}{tensor\hyp network}
\newacronym{CLIK-MPS}{CLIK-MPS}{complex-time Lindbladian Krylov subspace matrix\hyp product states}
\newacronym{LSE}{LSE}{local subspace expansion}
\newacronym{SVD}{SVD}{singular value decomposition}
\newacronym{QME}{QME}{quantum Mpemba effect}
\newacronym{KPZ}{KPZ}{Kardar\hyp Parisi\hyp Zhang}
\begin{document}
\author{Philipp Westhoff} \thanks{p.westhoff@physik.uni-muenchen.de}
\affiliation{Department of Physics, Arnold Sommerfeld Center for Theoretical Physics (ASC), Munich Center for Quantum Science and Technology (MCQST), Ludwig-Maximilians-Universit\"{a}t M\"{u}nchen, 80333 M\"{u}nchen, Germany}
\author{Sebastian Paeckel} \thanks{sebastian.paeckel@physik.uni-muenchen.de}
\affiliation{Department of Physics, Arnold Sommerfeld Center for Theoretical Physics (ASC), Munich Center for Quantum Science and Technology (MCQST), Ludwig-Maximilians-Universit\"{a}t M\"{u}nchen, 80333 M\"{u}nchen, Germany}
\author{Mattia Moroder}\thanks{moroderm@tcd.ie}
\affiliation{School of Physics, Trinity College Dublin, College Green, Dublin 2, D02K8N4, Ireland}
\def\thetitle{Fast and direct preparation of a genuine lattice BEC via the quantum Mpemba effect}
\title{\thetitle}
\begin{abstract}
%
%We present an efficient method for dissipatively preparing a \gls{BEC} directly on a lattice, avoiding the need for a two\hyp staged preparation procedure currently used in ultracold\hyp atom platforms.
%
\updated{We demonstrate that dissipative state preparation protocols in many\hyp body systems can be substantially accelerated via the quantum Mpemba effect.
Our approach exploits weak symmetries to analytically identify a class of simple, experimentally\hyp realizable states that converge exponentially faster to the steady state than typical random initializations.
In particular, we study the preparation of a lattice~\gls{BEC}, where the depletion can be controlled via the dissipation strength.
}
We also show how to tune the momentum of the created high\hyp fidelity \gls{BEC} by combining superfluid immersion with lattice shaking.
Our theoretical predictions are confirmed by numerical simulations of the dissipative dynamics.
This protocol paves the way to unlock the enormous potential of the dissipative preparation of highly
entangled states in analog quantum simulators.
\end{abstract}
\maketitle
Preparing and controlling highly\hyp entangled states is a central goal of analog quantum simulators based on ultracold atoms in optical lattices.
This includes, for instance, the realization of bosonic \cite{Greiner2002} and fermionic \cite{Jordens2008} Mott insulators, topological states \cite{Aidelsburger2013,Jotzu2014,Schweizer2019}, antiferromagnets \cite{Hart2015}, and many\hyp body localized states \cite{Schreiber2015}.
\updated{While traditional schemes are based on coherent control, it has recently been shown that adding controlled dissipation can be computationally~\cite{Zhan2025, Lin2025dissipativeprep} and experimentally~\cite{Mi2024} advantageous in terms of robustness to noise and efficiency.
%, where one exploits controlled dissipation,
Despite significant progress however, so\hyp called \gls{DSP} protocols still suffer from long preparation times~\cite{Pocklington2024Time-Ent-Tradeoff} and the problem of finding optimized, fast-converging initializations has remained completely unexplored. %(??divide the sentence in 2 and say that long times $\rightarrow$ many errors $\rightarrow$ many complex states are still beyond reach??)
}

\begin{figure}[H]
    \centering
    \subfloat[\label{subfig:bec:first:a}]{
        \includegraphics[width=0.95\columnwidth]{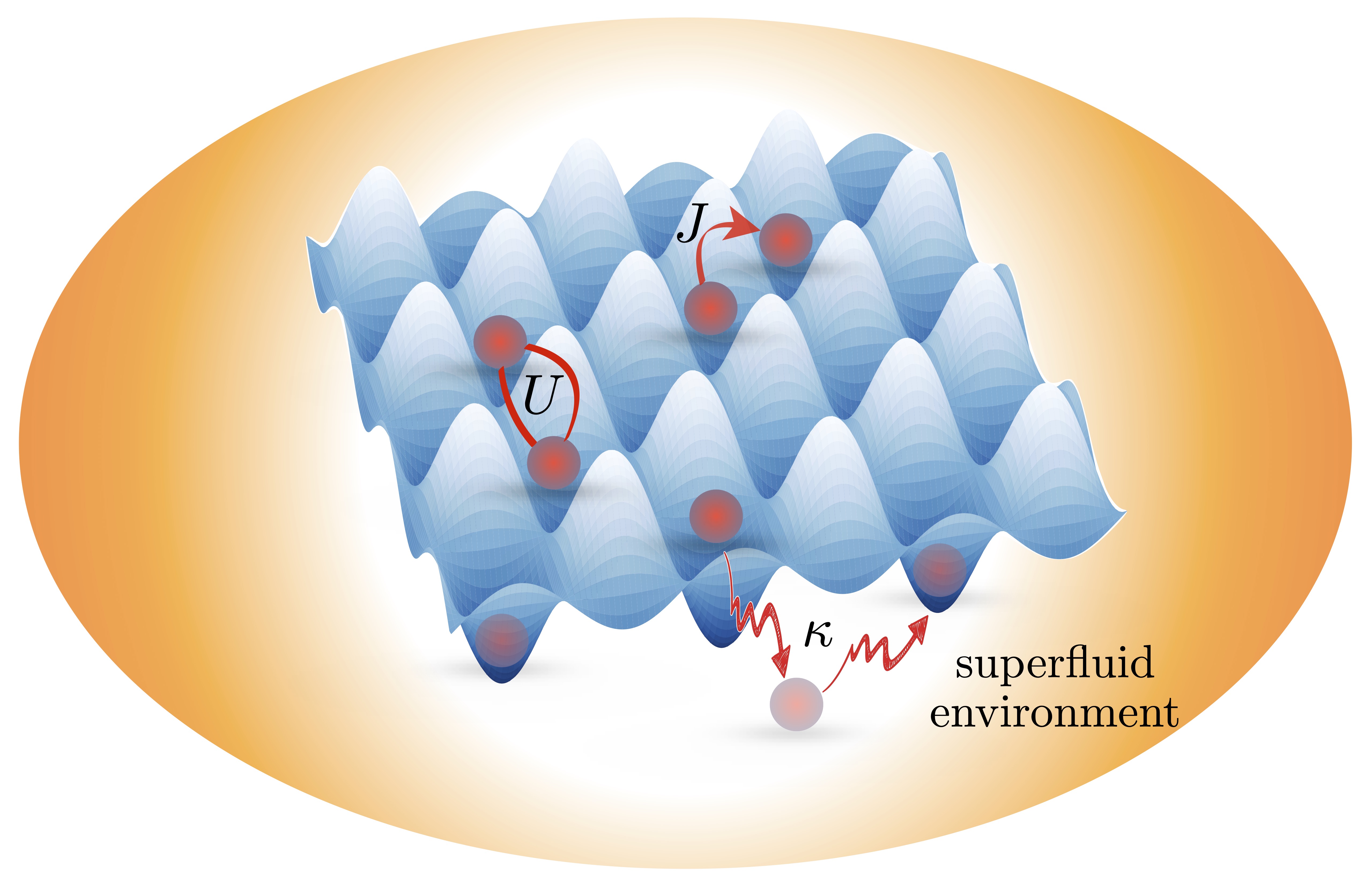}
    }
    \vspace{-0.4cm} 
    \subfloat[\label{subfig:bec:first:c}]{
        \includegraphics[width=0.95\columnwidth]{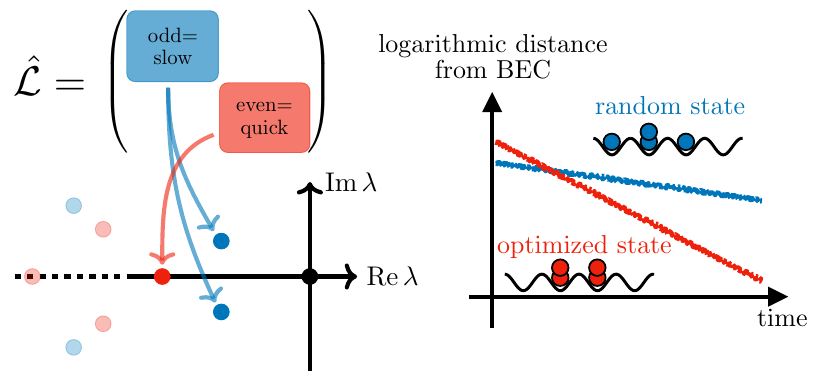}
    \hspace{-0.2cm}
    }
    \caption{
    Mpemba\hyp effect\hyp assisted preparation of a \acrfull{BEC} in an optical lattice.
    Panel a): A condensate of interacting bosonic particles can be prepared by combining coherent hopping with dissipation\hyp mediated tunneling via a superfluid environment in which the optical lattice is immersed.
    \updated{Panel b) left: we consider a Lindbladian $\hat{\mathcal{L}}$ featuring a weak $\mathds Z_2$ symmetry corresponding to reflections about the center of the lattice. 
    This endows the Lindbladian with a block-diagonal structure, with one block corresponding to evenly transforming states and one to oddly transforming states.
    Spectral analysis shows that the slowest-decaying mode belongs to the oddly transforming block.
    Panel b) right: Based on this symmetry argument, we identify a class of (evenly-transforming) product states that converge exponentially faster to the target \gls{BEC}.}
    }
    \label{fig:first}
\end{figure}

\updated{In this letter, we show that the Mpemba effect, which originally referred to the classical non\hyp equilibrium phenomenon of hot systems cooling faster than warm ones, can be exploited to substantially speed up \gls{DSP} protocols.
As a concrete many\hyp body example, we study the preparation of a~\acrfull{BEC} ~\cite{Diehl2008,Kraus2008} (see~\cref{subfig:bec:first:a}), and exploit a discrete symmetry of the system to identify simple, experimentally realizable initial configurations that converge exponentially faster to the steady state than random ones (see~\cref{subfig:bec:first:c}).
Note that the preparation of~\glspl{BEC} is of particular importance, since it represents the first step for preparing \textit{any} highly entangled state in ultracold atoms platforms.
Our analysis even carries over to finite\hyp momenta~\glspl{BEC} \cite{Tomadin2011}, which require the modulation of the environmentally\hyp mediated hopping by an additional complex phase via lattice shaking techniques~\cite{DiLiberto2011, Creffield2016,impertro2024}}.

\paragraph{The quantum Mpemba effect}
The dynamics of a quantum system weakly coupled to a Markovian (i.e. memoryless) environment obey the Lindblad master equation \cite{Lindblad1976}
\begin{equation}
    \frac{\mathrm{d} \hat{\rho}}{\mathrm{d}t} = \mathcal{L} \hat{\rho} = -\mathrm i [\hat{H}, \hat{\rho}] + \sum_l \hat{L}^\nodagger_l \hat{\rho} \hat{L}^\dagger_l - \frac{1}{2} \{ \hat{L}^\dagger_l \hat{L}^\nodagger_l, \hat{\rho}\}\;,
    \label{eq:Lindblad}
\end{equation}
where $[\cdot, \cdot]$ and $\{\cdot, \cdot\}$ indicate the commutator and anticommutator, respectively, $\hat{\rho}$ denotes the system's density matrix, $\hat{H}$ the Hamiltonian, and the influence of the environment is captured by so\hyp called jump operators $\hat{L}_l$.
In the eigenbasis of the Lindbladian superoperator $\mathcal{L}$, the time evolved\hyp density matrix can be written as \cite{Carollo2021}
\begin{equation}
    \hat{\rho}(t)=\mathrm e^{\mathcal{L}t}\hat{\rho}_0=\hat{\rho}_\mathrm{ss}+\sum_{p=2}^{D^2} \mathrm e^{\lambda_p t}\mathrm{Tr}(\hat{\rho}_0 \hat{l}_{p})\hat{r}_p \;.
    \label{eq:lindblad:decomposition}
\end{equation}
Here, $\hat{\rho}_0$ is the initial state, $\lambda_p$ indicates an eigenvalue of $\mathcal{L}$, $\hat{l}_{p}$ and $\hat{r}_{p}$ are its corresponding left and right eigenmodes and $D$ is the dimension of $\hat{H}$.
Since all $\lambda_p$ have negative real parts, every term in \cref{eq:lindblad:decomposition} decays exponentially in time except for the steady state $\hat{\rho}_\mathrm{ss}$, which, up to normalization, is given by the right eigenmode corresponding to the eigenvalue $\lambda_1$.
Let us sort the eigenvalues ascendingly according to the absolute value of their real part as $\lambda_1 = 0 < \abs{\Re(\lambda_2)} \leq \abs{\Re(\lambda_3)} \leq \cdots$.
At late times, a typical random state will approach the steady state with an equilibration speed $\propto \exp[\mathrm{Re}(\lambda_2)t]$.
Instead, special states that have zero overlap with the so\hyp called slowest decaying mode $\hat{l}_2$ will equilibrate exponentially faster, namely as $\exp[\mathrm{Re}(\lambda_3)t]$.
Given some distance function $\mathfrak{D}$ (for instance the $L_2$-norm) and assuming that initially the special, fast\hyp equilibrating states have a larger distance to the steady state $\hat \rho_\mathrm{ss}$ than typical, random ones, their distance curves w.r.t. $\hat \rho_\mathrm{ss}$ will cross as a function of time, which is called a Mpemba effect \cite{Lu2017}. %(as illustrated in~\cref{subfig:bec:first:c}).
\updated{Besides classical Markovian systems~\cite{Mpemba1969,Lu2017, Klich2019, Kumar2020, Kumar2022, Teza2025}, recently the Mpemba effect has been thoroughly investigated in isolated \cite{Ares2023Nat, Murciano2024Stat, Liu2024, Joshi2024, Rylands2024, Yamashika2024, Turkeshi2024} and open \cite{Nava2019, Carollo2021, Bao2022, Kochsiek2022, Ivander2023, Wang2024, Moroder2024, Strachan2024, Medina2024, Xu2025} quantum systems.
Yet, open many\hyp body systems remained elusive so far, because in general the decomposition~\cref{eq:lindblad:decomposition} can be computed only for small\hyp scale systems amenable to \gls{ED}.}

\paragraph{Preparing a lattice \gls{BEC}}
We consider bosonic particles in a \gls{1D} lattice described by the Bose\hyp Hubbard Hamiltonian
\begin{equation}
       \hat{H}_{k_0} = -J \sigma_{k_0} \sum_{j=1}^{L-1} \left( \mathrm e^{\mathrm ik_0 }\hat{b}^\dagger_{j+1} \hat{b}^\nodagger_{j} + \mathrm{h.c.} \right) + \frac{U}{2} \sum_{j=1}^{L} \big[\hat b^\dagger_j\big]^2 \big[\hat b^\nodagger_j\big]^2 \;.
    \label{eq:bose:hubbard:ham}
\end{equation}
Here, $\hat{b}^\dagger_j$ ($\hat{b}^\nodagger_j$) creates (annihilates) a boson on site $j$, $L$ is the number of sites, $J$ and $U$ represent the hopping amplitude and the onsite interaction strength, and we consider \glspl{OBC}.
The prefactor $\sigma_{k_0}$ is $1$ for $|k_0|<\pi/2$ and $-1$ otherwise, ensuring that the real part of the hopping amplitude is always negative for $J>0$ (we set the lattice spacing to one).
We study the Markovian dissipative dynamics obeying \cref{eq:Lindblad} and choose
\begin{equation}
    \hat{L}^{k_0}_j = \sqrt{\kappa}(\hat b^\dagger_{j+1} +  \mathrm e^{-\mathrm ik_0}\hat b^\dagger_j)(\hat b^\nodagger_{j+1} - \mathrm e^{\mathrm ik_0} \hat b^\nodagger_j)
    \label{eq:jump:op}
\end{equation}
as jump operators with dissipation strength $\kappa$.
Up to the phase factor $\mathrm e^{\mathrm ik_0}$ and boundary conditions, \cref{eq:bose:hubbard:ham,eq:jump:op} are consistent with the model proposed in \cite{Kraus2008, Diehl2008}.
These jump operators drive any initial state to a \gls{BEC} and can be experimentally realized by immersing the system in a superfluid \cite{Griessner2006}.
Using Bogoliubov theory \cite{Diehl2010, Tomadin2011}, we derive the steady states in each $k$\hyp sector, which are given by $\hat\rho^k_\mathrm{ss} = \mathrm e^{-\hat h_\mathrm{eff}^k/T_\mathrm{eff}}/Z_k$ (see the Supplemental Material~\cite{supp_mat_bec}).
Here $\hat h^k_\mathrm{eff} = E^k_\mathrm{eff} \hat a^\dagger_k\hat a^\nodagger_k$ is an effective single\hyp particle Hamiltonian, $T_\mathrm{eff}$ can be interpreted as the effective temperature and $Z_k = \mathrm{Tr}\big(\mathrm e^{-\hat h_\mathrm{eff}^k/T_\mathrm{eff}}\big)$.
The system's steady state can be written as $\hat \rho_\mathrm{ss}=\prod_k\hat \rho^k_\mathrm{ss}$.
Expanding $E^k_\mathrm{eff}$ in the limit $k\ll \sqrt{JU}/\kappa$, the effective temperature acquires a particularly simple form, which to the first non\hyp trivial order in the dissipation strength \updated{(which had not been considered in~\cite{Diehl2008})} is given by
\begin{equation}
    T_\mathrm{eff} = \frac{|U|n}{2\sqrt{1+(2n\kappa/J)^2}}\;,
    \label{eq:T:eff}
\end{equation}
 with the density $n=N/L$. Interestingly, in this limit, the effective $k$\hyp Hamiltonian also simplifies and reduces to~\cref{eq:bose:hubbard:ham}, when replacing the interaction $U\rightarrow U_\mathrm{eff} = U / (1+a^2)$, where $a=2n\kappa/J$.
We stress that these results include the case $U>\kappa$, while in the limit $\kappa \gg U$ and fixed $L$ we get $U_\mathrm{eff} \sim \mathcal {O}(U/\kappa)$ and $T_\mathrm{eff}=0$, as we discuss in the Supplemental Material \cite{supp_mat_bec}.
Note how dissipation suppresses the effective repulsive interaction, eventually generating a maximal condensation in the limit $\kappa/U\gg 1$.
This starkly contrasts the equilibrium case, where a true condensation is forbidden and underlines the non\hyp equilibrium character of the steady state.
In fact, our results suggest a picture in which lattice bosons scatter off the immersing~\gls{BEC}, yielding an effective pumping protocol.
As a consequence, the effective temperatures can be significantly lower than the one of the immersing~\gls{BEC}, and hence, the coherence of the prepared lattice~\gls{BEC} much higher.
To illustrate this point, we perturbatively evaluated the fraction of non\hyp condensed particles, known as condensate depletion, finding $\delta = (N-N_0)/N \sim \mathcal O((U/\kappa)^2)$, where $N_0$ denotes the occupation of the condensate. Similarly, for the two\hyp point correlation functions we obtain $\langle \hat{b}^\dagger_i \hat{b}^\nodagger_j \rangle \stackrel{i-j=L}{\longrightarrow}n-\mathcal{O}(U/\kappa)^2$, which we denote as a lattice analog of off\hyp diagonal long\hyp range order~\cite{Penrose1956,Yang1962, Tindall2019}.
Note that both quantities can be controlled by tuning $U/\kappa$.
Strikingly, the condensate depletion vanishes as $(U/\kappa)^2$, which suggests that \glspl{BEC} with an extremely large condensate fraction can be realized.
Finally, using the phase modulation $\mathrm e^{\mathrm ik_0}$, which can be implemented via lattice shaking \cite{Creffield2016,impertro2024}, i.e. by adding a fast\hyp oscillating laser field, the mode $k_0$ in which the particles condense can be tuned, allowing for the fast realization of high\hyp quality finite\hyp momentum \glspl{BEC}.
We note that the case $k_0=\pi$ is particularly simple to realize experimentally, since $\mathrm e^{\mathrm i\pi} = -1$ implies that both the coherent and the incoherent hopping are real and thus no lattice shaking is required.
\begin{figure}[H]
    \centering
    \subfloat[\label{subfig:BEC:a}]{
        \includegraphics[width=\columnwidth]{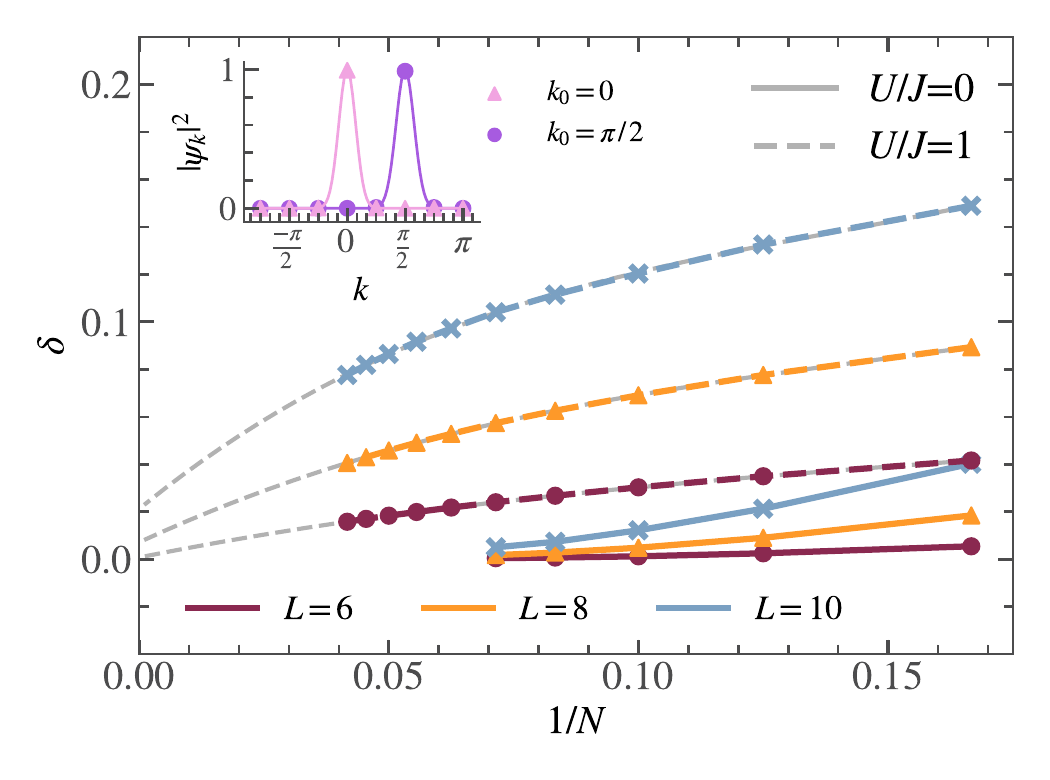}
    }
    \vspace{-2em} % Add some vertical space between subfigures
    % Panel (c)
    \subfloat[\label{subfig:BEC:b}]{
        \includegraphics[width=0.8\columnwidth]{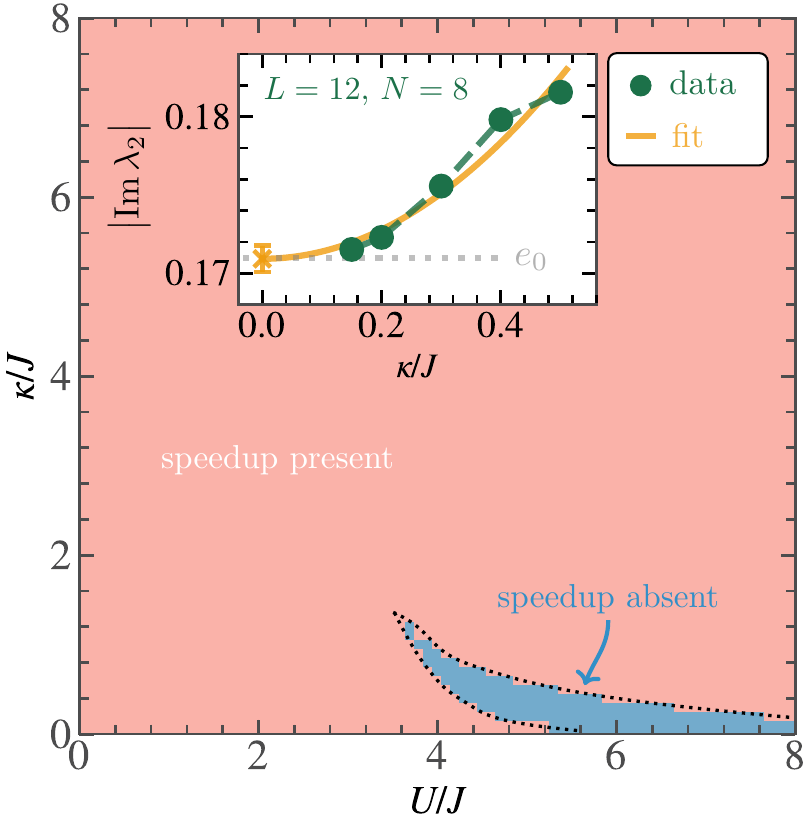}
    }
    \caption{
        Panel a): Simulating the dissipative preparation of a finite\hyp momentum \gls{BEC} in a \gls{1D} lattice.
        We show the condensate depletion as a function of the inverse total particle number $N$ for $|k_0|<\pi/2$.
        %
        %\discussive{The extensive scaling is clearly visible for $L=6$ and $L=8$, while for $L=10$ larger particle numbers would be required.} 
        The extrapolations to $1/N \to 0$ (gray lines) were performed with a third\hyp order polynomial in $1/N$.
        Inset: Momentum space representation of the eigenvector $\psi$ corresponding to the eigenvalue $N_0$ of $\bm\gamma$ for characteristic Lindbladian momenta $k_0=0$ and $k_0=\nicefrac{\pi}{2}$ (see \cref{eq:jump:op}).
        \updated{Panel b): parameter regions in which symmetric initial states~\cref{eq:symmetric:states} yield a Mpemba speedup (red) and in which they don't (blue). These two regions correspond to the slowest\hyp decaying mode of the Lindbladian transforming oddly and evenly under the inversion symmetry~\cref{eq:inversion:symmetry}, respectively, which originate from a level crossing \cite{Teza2023, Yaacoby2024, Teza2023Relaxaion}. We consider $L=6$ and $N=3$ at $k_0<\pi/2$ and employ \gls{ED}. Inset: Criterion to numerically validate the occurrence of the Mpemba effect. In the limit $\kappa\to0$, the imaginary part of $\lambda_2$ (green dots) converges to the eigenvalue $e_0$ of the first excited eigenspace of $\hat{\mathcal{H}}$ in perturbation theory (see~\cref{eq:vectorized:unitary:part:of:L}). For the extrapolation we employ a quadratic fit (orange line) and consider $U=0$.} 
    }
    \label{fig:BEC}
\end{figure}
\paragraph{Symmetry\hyp adapted initial states}
A set of fast\hyp converging initial states can be found without explicitly diagonalizing $\mathcal{L}$.
For that purpose, we exploit the fact that the Lindbladian and its unique steady state are invariant under a (discrete) symmetry.
We emphasize that here and in the following, we refer to symmetry transformations on the vectorized, i.e., doubled, Hilbert space, which should not be confused with transformations on the physical Hilbert space only.
Specifically, for $k_0=0$ the Lindbladian constructed from~\cref{eq:bose:hubbard:ham,eq:jump:op} possesses the same inversion symmetry as the Hamiltonian~\cref{eq:bose:hubbard:ham}, i.e. it is invariant under reflections about the center of the lattice described by the unitary transformation
\begin{equation}
\hat{U}^\nodagger_\mathrm{inv}\hat b^\dagger_{j}\hat{U}_\mathrm{inv}^\dagger = \hat b^\dagger_{L+1-j} \;.
\label{eq:inversion:symmetry}
\end{equation}
The more general case is covered in Supplemental Material~\cite{supp_mat_bec}.
This weak symmetry \cite{Buča_2012, Albert2014} decomposes $\mathcal{L}$ into two blocks, corresponding to eigenmodes that transform evenly or oddly under $\hat{U}_\mathrm{inv}$.
We investigate the unitary part of the Lindbladian $\mathcal{H}$ (see \cref{eq:Lindblad}), which in its vectorized form is given by~\cite{Landi2022} 
\begin{equation}
    \hat{\mathcal{H}} = -\mathrm i \hat{H} \otimes  \hat{\mathds{1}} + \hat{\mathds{1}} \otimes \mathrm i\hat{H}^{\scriptscriptstyle T} \; .
    \label{eq:vectorized:unitary:part:of:L}
\end{equation}

In the Supplemental Material~\cite{supp_mat_bec}, we show that its vectorized eigenstates are adiabatically connected to those of $\mathcal{L}$ using perturbation theory in the limit $\kappa \to 0^+$, \updated{which is also confirmed using large\hyp scale numerics \cite{westhoff2025tensor}}.
We furthermore find that the slowest\hyp decaying mode $\hat{l}_2$ transforms oddly under inversion \updated{for almost all parameters $(U,\kappa)$, as we discuss below}.
An important consequence of the previous considerations is that any physically\hyp realizable state that is symmetric under reflections about the center of the lattice has zero overlap with $\hat{l}_2$ and equilibrates exponentially faster to the \gls{BEC} than random initial states.
This also includes the product states
\begin{equation}
    \ket{\psi} = \ket{n_1, n_2, \dots, n_{\nicefrac{L}{2}}, n_{\nicefrac{L}{2}}, \dots n_2, n_1} \;,
    \label{eq:symmetric:states}
\end{equation}
where the $j$-th entry indicates the number of particles on site $j$.
Common examples of states of the type \cref{eq:symmetric:states} include so\hyp called wedding\hyp cake states $\ket{\psi} = \ket{1,\,2,\dots, \nicefrac{L}{2}-1,\, \nicefrac{L}{2}, \, \nicefrac{L}{2}, \, \nicefrac{L}{2}-1, \dots,2, \, 1}$, which can be readily prepared in harmonic traps.
Among such symmetric, fast\hyp converging product states, the fastest\hyp converging is the one where all particles are initially located on the central site(s) (see \cref{subfig:bec:first:c}).
This can be qualitatively understood from the fact that such a state is connected to the \gls{BEC} by a minimal number of hoppings and we call it the \gls{SL} state.
Crucially, the \gls{SL} state can also be realized experimentally in the novel hybrid setups combining optical lattices with optical tweezers \cite{Young2022,Renhao2024}.
However, we want to point out that all states of the form~\cref{eq:symmetric:states} exhibit exponential speedups.
Note that the same symmetry arguments can be applied to higher\hyp dimensional systems, and preliminary numerical evidence shows that the \gls{SL} state is the fastest\hyp converging one also in 2D.
The arguments outlined above directly apply to the zero momentum case $k_0=0$, only.
However, Lindbladians with different characteristic momenta $k_0$ (see \cref{eq:jump:op}) and their eigenmodes are unitarily connected to one another, as we discuss in the the Supplemental Material \cite{supp_mat_bec}.

\paragraph{Simulating the dissipative dynamics}
\updated{To quantitatively assess the speed and the accuracy of our protocol}, we employ \glspl{MPS} techniques to numerically compute the dissipative evolution generated by \cref{eq:bose:hubbard:ham,eq:jump:op} for \gls{1D} many\hyp body systems.
We represent vectorized density matrices as \glspl{MPS} \cite{Schollwoeck2011, Wolff2020} and vectorized Lindbladians as \glspl{MPO} \cite{Prosen2009, Casagrande2021, Somoza2019, Moroder_diss}.
The time evolution is computed with a variant of the \gls{TDVP} \cite{Haegeman2011,Yang2020,Grundner2023,Paeckel2019} tailored for bosonic systems (see the Supplemental Material \cite{supp_mat_bec} for the details of the numerical implementation).

First, in \cref{subfig:BEC:a} we show the accuracy of the obtained \glspl{BEC} at different system sizes.
For this purpose, we compute the condensate depletion $\delta$ as the deviation of the leading eigenvalue $N_0$ of the one\hyp body density matrix $\bm\gamma$,  componentwise defined via $\bm\gamma_{ij} = \langle \hat{b}^\dagger_j\hat{b}^\nodagger_i\rangle$, from the maximal possible condensate population $N$.
The condensate depletion decreases upon increasing the total particle number $N$ and we find an extensive scaling $\delta \sim \mathcal O(1/N)$, in agreement with the Bogoliubov theory.
Strikingly, extrapolating the condensate depletion as a function of $1/N$ to the limit $N\rightarrow \infty$, we observe $\delta \rightarrow 0$ also at finite interaction strengths $U/J=1$ (dashed lines).
Moreover, the inset indicates that the eigenvector corresponding to the leading eigenvalue of $\bm\gamma$ has almost unit weight at the characteristic momentum of the Lindbladian $k_0$, demonstrating the formation of a finite\hyp momentum~\gls{BEC}, controlled by $k_0$.

\updated{Then, in \cref{subfig:BEC:b} we show that in the largest part of the parameter space, the symmetric initial states~\cref{eq:symmetric:states} yield a Mpemba speedup.
These \gls{ED} results are supported by large\hyp scale \gls{MPS} results in the Supplemental Material~\cite{supp_mat_bec}, which demonstrate that the small area in which the Mpemba speedup is absent does not grow upon increasing the system size.
In the inset we plot $|\Im\lambda_2|$ as a function of $\kappa$, which shows the slowest decaying mode of the Lindbladian stems from the first excited eigenspace of $\hat{\mathcal{H}}$ (see~\cref{eq:vectorized:unitary:part:of:L}), whose eigenvalue we denote as $\mathrm ie_0$, validating the perturbation theory approach that we also analyze in detail in the Supplemental Material~\cite{supp_mat_bec}.
}

\begin{figure}[h]
    \centering
    \subfloat[\label{subfig:speedups:a}]{
        \includegraphics[width=\columnwidth]{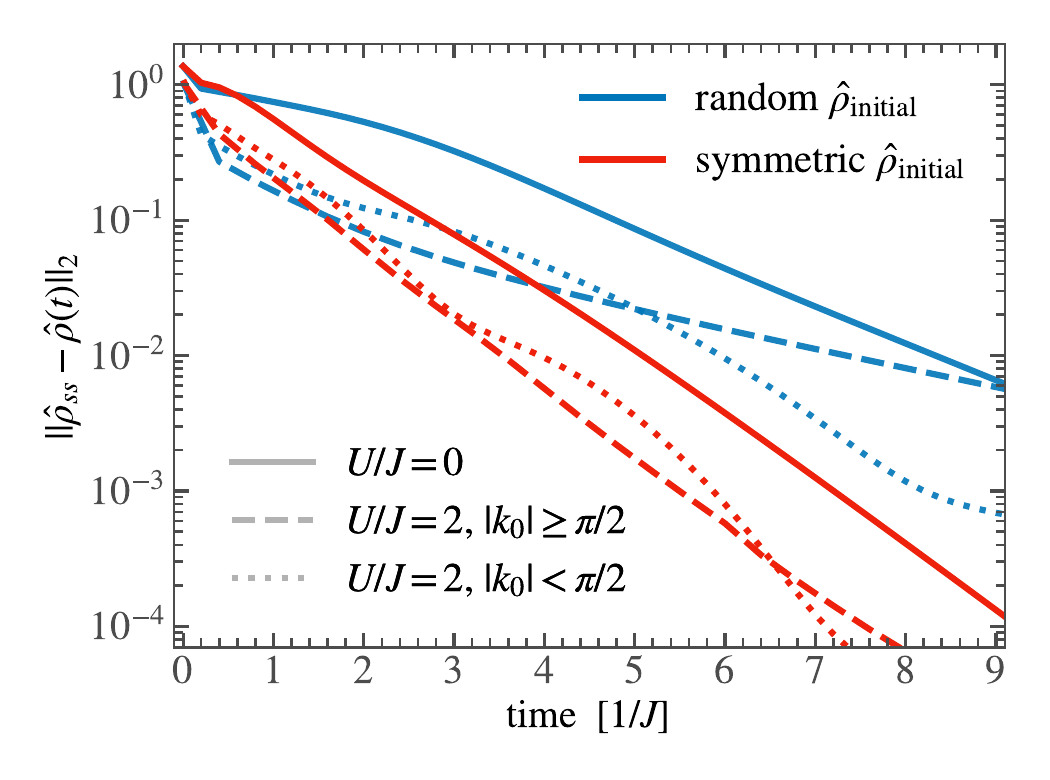}
    }
    \vspace{-3.5em} 
    % Panel (c)
    \subfloat[\label{subfig:speedups:b}]{
        \includegraphics[width=\columnwidth]{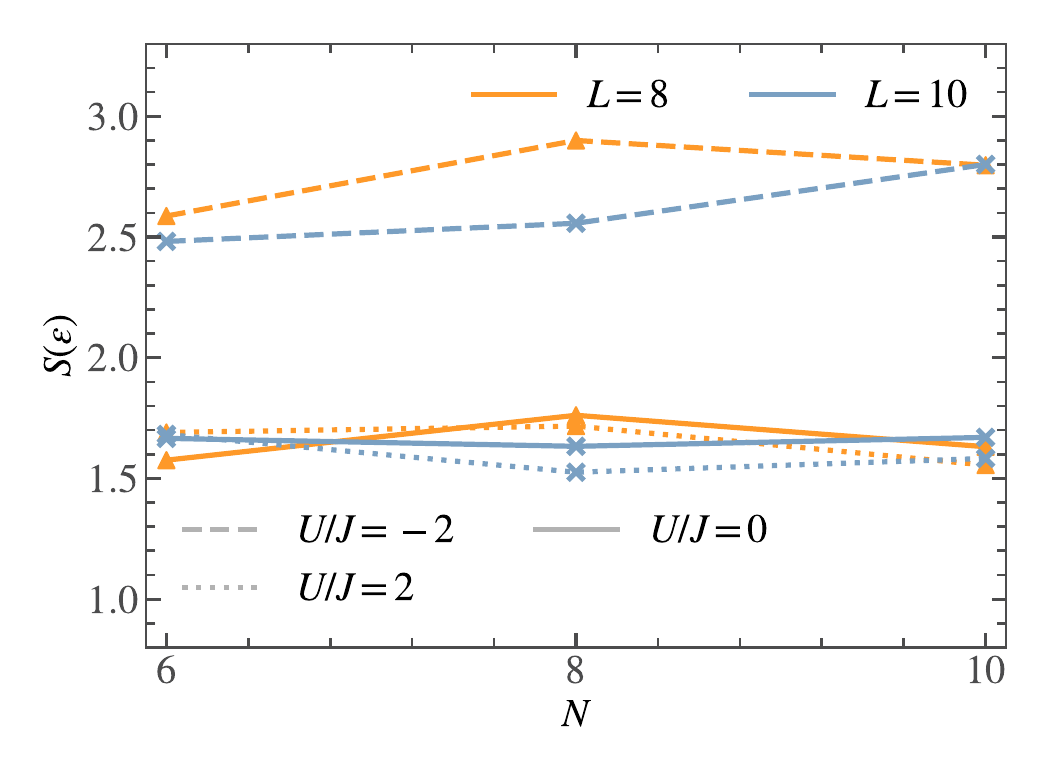}
    }
    \caption{
        Mpemba\hyp speedups in the preparation of \glspl{BEC}.
        Panel a): The \acrfull{SL} state (red lines), where all particles are initially located on the central site(s), converges exponentially faster to the steady state $\hat \rho_\mathrm{ss}$ than random initial product states (blue lines).
        Random states are generated by distributing $N$ particles on the lattice, sampling the positions from a uniform distribution over the sites.
        Line styles indicate different bosonic interaction strengths and characteristic momenta $k_0$.
        All calculations were performed with system parameters $L=10$, $N=10$, local dimension $d=N+1$, and $\kappa=2J$, and we averaged over $5$ product state realizations.
        Panel b): The corresponding speedups $S(\epsilon) = t_\mathrm{random}/t_\mathrm{symmetric}$ as a function of the total particle number. \updated{Note that the speedup is dictated by the spectrum of the Lindbladian, namely $S(\epsilon) \to \mathrm{Re}\, \lambda_3/\mathrm{Re}\, \lambda_2$ as $\epsilon\to 0$}.
        $t_\mathrm{random}$ and $t_\mathrm{symmetric}$ are the times at which the $L_2$-distance from the steady state has dropped below a precision threshold $\epsilon=10^{-4}$ for the random and the \gls{SL} states, respectively.
    }
    \label{fig:speedups}
\end{figure}

Next, we quantify the Mpemba speedups yielded by our protocol, simulating a lattice loaded with either the~\gls{SL} state or random product states and different values of $U/J$.
In \cref{subfig:speedups:a}, we compare the equilibration dynamics towards the \gls{BEC}. 
As a distance measure, we consider the $L_2$\hyp norm $\mathfrak{D}(t)=\lVert\hat{\rho}(t)-\hat{\rho}_\mathrm{ss}\rVert_2$, which is the easiest to compute in the vectorized framework.
We find that the \gls{SL} state converges exponentially faster than random product states for all considered bosonic repulsion strengths $U/J$. 
Taking a more practical perspective, the actual speedup to achieve a certain condensate fidelity is the relevant quantity.
For that purpose, \cref{subfig:speedups:b} shows the corresponding speedups $S_\epsilon$ relative to a desired $L_2$\hyp distance $\epsilon$ from the steady state $\hat{\rho}_\mathrm{ss}$.
Notably, the speedups do not change significantly when varying the system size or the number of particles. 
This indicates that similar speedups can also be achieved for larger system sizes.
Note that due to unitary equivalence, the speedups are the same for all $|k_0| <\pi/2$ and for all $|k_0| \geq \pi/2$ \cite{supp_mat_bec}.

\paragraph{Discussion and conclusions} 
\updated{\Acrfull{DSP} is finally becoming a viable experimental alternative to coherent control\hyp based preparation schemes in terms of robustness and efficiency~\cite{Mi2024, Lloyd2025PRXQuantum, Zhan2025, Lin2025dissipativeprep}, but still suffers from long preparation times.
In this letter, we show that the Mpemba effect can be exploited to find simple, experimentally\hyp realizable initializations that substantially reduce the protocol times. 
In particular, we study the important example of a~\gls{BEC}~\cite{Diehl2008,Kraus2008}, which represents the starting point for preparing any highly entangled state in ultracold atoms.
Crucially, we exploit a weak symmetry of the system to find optimized initializations without diagonalizing the Lindbladian, which allows us to study a genuine many\hyp body system.
In the limit of strong dissipation, we analytically show that the steady state exhibits true long\hyp range order on the length scales of the lattice.
Moreover, we use \gls{MPS} to simulate the protocol numerically and quantitatively estimate the obtained speedups.
The numerical results also confirm the predicted scaling for the depletion $\delta \propto (U/\kappa)^2$ already for moderate dissipation strength.
}

\updated{Our symmetry\hyp based approach to find fast\hyp converging initial states can be applied to a wide class of many\hyp body open quantum systems.
This significantly reduces the preparation times and the associated errors of \gls{DSP} protocols, paving the way to the realization of complex states, such as $\eta$\hyp paired superconducting states in ultracold atoms~\cite{Tindall2019}, which have thus far remained elusive.}

\paragraph{Acknowledgements}
We are grateful to Monika Aidelsburger for enlightening discussions and  Henning Schlömer, Timothy Harris and Federico Carollo for useful comments on a previous version of the manuscript.
PW and SP acknowledge support by the Deutsche Forschungsgemeinschaft (DFG, German Research Foundation) under Germany’s Excellence Strategy-426 EXC-2111-390814868.
This work was supported by Grant No. INST 86/1885-1 FUGG of the German Research Foundation (DFG).
MM acknowledges funding from the Royal Society and Research Ireland.
All calculations were performed using the \textsc{SyTen} toolkit \cite{hubig:_syten_toolk,hubig17:_symmet_protec_tensor_networ}.
%de

%\bibliographystyle{naturemag}  % Mimics Nature-style formatting
\bibliography{Literature}
\end{document}

\author{Philipp Westhoff} \thanks{p.westhoff@physik.uni-muenchen.de}
\affiliation{Department of Physics, Arnold Sommerfeld Center for Theoretical Physics (ASC), Munich Center for Quantum Science and Technology (MCQST), Ludwig-Maximilians-Universit\"{a}t M\"{u}nchen, 80333 M\"{u}nchen, Germany}
\author{Sebastian Paeckel}\thanks{sebastian.paeckel@physik.uni-muenchen.de}
\affiliation{Department of Physics, Arnold Sommerfeld Center for Theoretical Physics (ASC), Munich Center for Quantum Science and Technology (MCQST), Ludwig-Maximilians-Universit\"{a}t M\"{u}nchen, 80333 M\"{u}nchen, Germany}
\author{Mattia Moroder}\thanks{moroderm@tcd.ie}
\affiliation{School of Physics, Trinity College Dublin, College Green, Dublin 2, D02K8N4, Ireland}
%\affiliation{Trinity Quantum Alliance, Unit 16, Trinity Technology and Enterprise Centre, Pearse Street, Dublin 2, D02YN67, Ireland}
%
\def\thetitle{Supplemental Material for \\``Fast and direct preparation of a genuine lattice BEC via the quantum Mpemba effect''}
\title{\thetitle}
\maketitle

\section{Exponentially faster converging product states} \label{sec:app:fast:states}
In this section, we prove that the symmetric states considered in the main text converge exponentially faster to the steady state than random states. Here, we first consider the special case of the zero\hyp momentum condensate. Then, in \cref{sec:app:unitary:equivalence} we generalize this result to finite momenta.

Consider the Bose Hubbard Hamiltonian at $k_0=0$
\begin{equation}
    \hat{H} = -J \sum_{j=1}^{L-1} \left( \hat{b}^\dagger_{j+1} \hat{b}_j^\nodagger + \mathrm{h.c.} \right) + \frac{U}{2} \sum_{j=1}^{L} \big[\hat{b}^\dagger_j \big]^2\big[\hat{b}^\nodagger_j\big]^2 \;,
    \label{eq:app:bose:hubbard:ham}
\end{equation}
with \glspl{OBC} that we analyzed in the main text.
It features a discrete symmetry that we label \textit{inversion symmetry}, represented by the unitary $\hat{U}_\text{inv}$ with action
\begin{equation}
    \hat{U}_\text{inv}^\nodagger\hat{b}^\dagger_{j}\hat{U}_\text{inv}^\dagger = \hat{b}^\dagger_{L+1-j}\;.
    \label{eq:app:inversion:symmetry}
\end{equation}
In the case of vanishing interaction $U=0$, the transformation
\begin{align}
    \hat{d}^\dagger_k = \sum_{j=1}^L \sin (kj) \; \hat{b}^\dagger_j\, , \quad k(m) = \frac{\pi}{L+1}m, \; m\in \{ 1, \ldots, L\}
\end{align}
brings the Hamiltonian into the diagonal form
\begin{equation}
    \hat{H} = \sum_{k} E_k \hat{d}^\dagger_k\hat{d}^\nodagger_k, \quad \quad E_k = -2J\cos(k) \; .
    \label{eq:app:diagonalized:H}
\end{equation}
The eigenvectors are constructed by successively adding particles $\hat{d}^\dagger_k$ into the system.
They are characterized by a vector $\mathbf{n} = (n_1, n_2, \dots, n_L)$, which describes the occupation of each of the modes $m$,
\begin{equation}
    \ket{\mathbf{n}} = \prod_{m=1}^L \frac{(\hat{d}^\dagger_{k(m)})^{n_m}}{\sqrt{n_m!}}\ket{0} \; .
\end{equation}
The corresponding eigenvalues read
\begin{equation}
 E(\mathbf{n}) = \sum_{m=1}^LE_{k(m)} n_m \;,
\end{equation}
and obey
\begin{equation}
 E_{k(m)} =- E_{k(L+1-m)} \;.
    \label{eq:app:eigenvalues:property}
\end{equation}
Moreover, the eigenvectors are also eigenvectors of the inversion symmetry $\hat{U}_{\text{inv}}$, and their eigenvalue is alternating, i.e
\begin{equation}
    \hat{U}_\text{inv}^\nodagger\hat{d}^\dagger_{k(m)}\hat{U}_\text{inv}^\dagger = (-1)^{m-1}\hat{d}^\dagger_{k(m)} \;.
    \label{eq:app:eigenvectors:property}
\end{equation}
Note that \cref{eq:app:eigenvalues:property} stems from the cosine in the eigenvalues and \cref{eq:app:eigenvectors:property} from properties of the sine. 
This implies that
\begin{equation}
    \hat{U}_{\text{inv}}\ket{\mathbf{n}} = \ket{\mathbf{n}} \prod_{m=1}^L (-1)^{n_m(m-1)} \, ,
    \label{eq:app:trafo:behavior}
\end{equation}
which reduces calculating the transformation behavior of a many\hyp body state to counting the number of particles in each mode and then multiplying all their eigenvalues.\\

\begin{figure}
    \centering
    \subfloat[\label{fig:suppmat:a}]{
    \includegraphics[width=0.8\columnwidth]{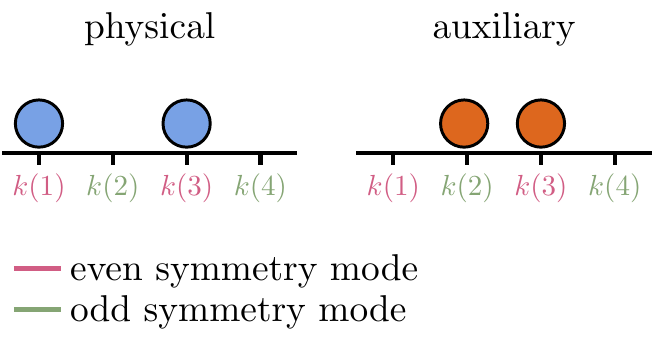}
 } 
    \vspace{0.5em}
    \subfloat[\label{fig:suppmat:b}]{
    \includegraphics[width=0.8\columnwidth]{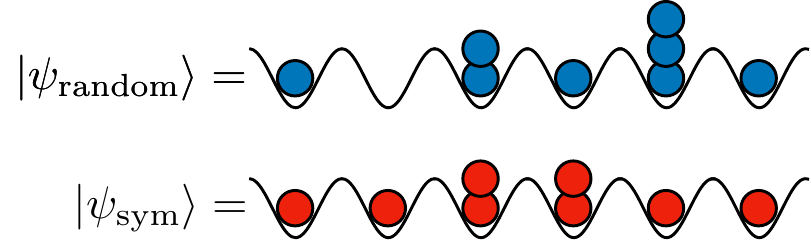}
 }
\caption{Panel a): A state in the eigenspace of $\hat{\mathcal{H}}$ to eigenvalue $-\mathrm i(E_{k(1)}-E_{k(2)})$ for an even ($L=4$) number of sites $L$. Notice, that the state transforms oddly, easily seen by the transformation behavior of every individual particle and \cref{eq:app:trafo:behavior}. Panel b): a random (blue) and symmetric (red) state on the real lattice. The random state has no fixed transformation behavior. The symmetric state transforms evenly, and hence athe corresponding density matrix $\ket{\psi_\mathrm{sym}}\hspace{-0.9mm}\bra{\psi_\mathrm{sym}}$ also transforms evenly.
}
\label{fig:physical:and:auxiliary:states}
\end{figure}
In order to study the Lindbladian, it is necessary to consider the enlarged, vectorized Hilbert space.
The general form of the vectorized Lindbladian \cite{Landi2022} is given by 
\begin{equation}
    \begin{aligned}
       &\hat{\mathcal{L}} = \underbrace{-\mathrm i \hat{H} \otimes  \hat{\mathds{1}} +   \hat{\mathds{1}} \otimes \mathrm i\hat{H}^{\scriptscriptstyle T} }_{\hat{\mathcal{H}}} + \\
       & \underbrace{\sum_l \hat{L}^{\nodagger}_l \otimes \left ( \hat{L}^{\dagger}_l \right ) ^{\scriptscriptstyle T} -\frac{1}{2} \hat{L}^{\dagger}_l \hat{L}^{\nodagger}_l \otimes  \hat{\mathds{1}}  - \frac{1}{2}  \hat{\mathds{1}} \otimes \left( \hat{L}^{\dagger}_l \hat{L}^{\nodagger}_l \right) ^{\scriptscriptstyle T}}_{\hat{\mathcal{D}}}.
    \end{aligned}
    \label{eq:app:Lindblad:vectorized}
\end{equation} 
For $k_0=0$, the jump operators are
\begin{equation}
    \hat{L}_j = \sqrt{\kappa}(\hat{b}^\dagger_{j+1} + \hat{b}^\dagger_{j})(\hat{b}_{j+1}^\nodagger - \hat{b}_j^\nodagger) \;,
    \label{eq:app:jump:ops}
\end{equation}
where $\kappa$ is the dissipation strength and $j \in \{1, \ldots,L\}$.
In the following, we focus on the unitary part of the Lindbladian $\hat{\mathcal{H}}$, which satisfies  
\begin{equation}
    \hat{\mathcal{H}} \,\ket{\mathbf{n}}\otimes\ket{\tilde{\mathbf{n}}} = -\mathrm i\big{(}E(\mathbf{n})-E(\tilde{\mathbf{n}})\big{)}\ket{\mathbf{n}}\otimes\ket{\tilde{\mathbf{n}}}\; .
    \label{eq:app:eigenvals:vectorized:H}
\end{equation}
For clarity, we will label the quantum number on the physical lattice $k_p$ and on the auxiliary $k_a$.
Notice, however, that the physical states in this enlarged Hilbert space have to satisfy the constraint of equal total particle number on both the physical and auxiliary lattices.
Thus, eigenvectors obey
 $\lVert \mathbf{n}\rVert_1 = \lVert \tilde{\mathbf{n}}\rVert_1$. The generalization of the inversion symmetry to the vectorized space reads $\hat{\mathcal{U}}_\text{inv}^\nodagger = \hat{U}_\text{inv}^\nodagger\otimes \hat{U}_\text{inv}^{*}$.
 We depict examples of odd vectorized states in \cref{fig:physical:and:auxiliary:states}.
 With these preliminary considerations at hand, we can prove the following important observation about a certain eigenspace of $\hat{\mathcal{H}}$.\\

 \textit{1. Theorem}: Let $\vecket{\psi}$ be an eigenstate of $\hat{\mathcal{H}}$ to eigenvalue $-\mathrm i(E_{k(1)}-E_{k(2)})$. Then it transforms oddly under inversion, that is $\hat{\mathcal{U}}_\text{inv}^\nodagger\vecket{\psi} = -\vecket{\psi}$.\\

 If this theorem is proven, we know that symmetric states do not have any overlap with this particular eigenspace.\\

\textit{Proof of 1}: We may take a general ket in the subspace with eigenvalue $-\mathrm i(E_{k(1)}-E_{k(2)})$, characterized by $\mathbf{n}$ and $\tilde{\mathbf{n}}$. This eigenvalue can only be obtained if all but two particles contribute a sum of 0 to the eigenvalue, while the remaining two have exactly the eigenvalue of interest (see \cref{eq:app:eigenvals:vectorized:H}).
Thus, we can reduce to the case with two physical and two auxiliary particles, but additionally we need to prove, that states with eigenvalue 0 transform evenly (proven in theorem 2). Lets first focus on the former.\\
For two particles (that means two physical and two auxiliary ones), two particles need to create the energy difference. Consequently, the possible fillings are: $k_p(1)$ and $ k_p(L-1)$; $k_a(L)$ and $k_a(2)$; $k_p(1)$ and $k_a(2)$; $k_p(L-1)$ and $k_a(L)$ (the last two stem from the special symmetry of the dispersion relation). In the case of even $L$, the first two transform evenly, the last two oddly. For odd $L$, all transform oddly. The two remaining particles combined need to have energy 0. If this is achieved by one particle on the physical and one on the auxiliary lattice, the two particles need to occupy $k_a(m)$ and $k_p(m)$, and thus transform evenly. If they both sit in one lattice, they need to occupy $k_{p/a}(m)$ and $k_{p/a}(L+1-m)$ and together transform oddly in the case of even $L$, and evenly for odd $L$. In total, all possibilities transform oddly according to \cref{eq:app:trafo:behavior}.\\

\noindent To conclude the proof, we need the following theorem.\\

 \textit{2. Theorem}: Let $\vecket{\psi}$ be an eigenstate of $\hat{\mathcal{H}}$ to eigenvalue 0. Then it transforms evenly under inversion, that is, $\hat{\mathcal{U}}_\text{inv}^\nodagger\vecket{\psi} = \vecket{\psi}$.\\

 \textit{Proof of 2}: We will first reduce the general particle number case to the case with one or two particles. The eigenvalue can be rewritten as 
\begin{align}
 E(\mathbf{n}) - E(\tilde{\mathbf{n}}) = \sum_{m=1}^{\lfloor\frac{L}{2}\rfloor} \big{(} n_m-\tilde{n}_m - n_{L+1-m} + \tilde{n}_{L+1-m} \big{)}E_{k(m)} \; .
\end{align}
The left side vanishes under the condition
\begin{equation}
 n_m-\tilde{n}_m - n_{L+1-m} + \tilde{n}_{L+1-m}=0, \quad \forall \, m\leq \Big{\lfloor} \frac{L}{2}\Big{\rfloor} \; .
\end{equation}
In case of odd $L$ care must be taken, as $n_{\nicefrac{L+1}{2}}-\tilde{n}_{\nicefrac{L+1}{2}}$ might be nonzero, as $E_{\nicefrac{L+1}{2}}=0$ already.
From now on, the $k$-sites $m$ and $L+1-m$ on the physical and auxiliary lattice together will be called $m$-th \textit{sector}.
In each of the sectors, the constraints must be fulfilled (with the exception of the \nicefrac{L+1}{2}\hyp sector, which we will refer to as \textit{zero\hyp sector}).\\
This also entails an \textit{even} number of particles in each sector. Since the energies are incommensurable, there always exist two particles that contribute a combined eigenvalue of $0$.
So, either one is on the physical lattice and one on the auxiliary lattice, or both are on the same lattice. In the latter case, we also need two particles on the other lattice that have an eigenvalue of zero. 
Taking away these $2$ or $4$ particles, we end up with a physical state with $N-1$ or $N-2$ particles that has the same eigenvalue.
The $2$ or $4$ particles taken out also had eigenvalue $0$, so by lemma 3 below, it transforms evenly. 
Doing this successively until no particles are left, the problem is reduced to the two and one\hyp particle case, which is the content of the next Lemma.\\

\textit{3. Lemma}: Theorem 2 holds in the case of one and two (physical) particles.\\

\textit{Proof of 3}:
We will make a case distinction between even and odd number of sites $L$. Starting with an \textit{even} number of sites and one (physical) particle, to fulfill the subspace constraint, the physical and the auxiliary particle need to sit in the same sector, denoted by $m$. It reads
\begin{equation}
 n_m + \tilde{n}_{L+1-m} = n_{L+1-m} + \tilde{n}_m = 1 \, ,
\end{equation}
giving the possibilities $n_m=\tilde{n}_m=1$ or $n_{L+1-m}=\tilde{n}_{L+1-m} = 1$, and both possible configurations transform evenly.\\
For two physical particles, there are more possible fillings. Either all $4$ particles ($2$ physical and $2$ auxiliary) are loaded in one sector, or two sectors are filled with two particles each. In the former case, the subspace constraint enforces
\begin{equation}
 n_m + \tilde{n}_{L+1-m} = n_{L+1-m} + \tilde{n}_m = 2
\end{equation}
for one sector $m$, which gives rise to the fillings $n_m=\tilde{n}_m=2$, $n_{L+1-m}+\tilde{n}_{L+1-m} = 2$, $n_m = \tilde{n}_m=n_{L+1-m}=\tilde{n}_{L+1-m} = 1$, all transforming evenly. 
Next, there is also the possibility of just two particles sitting in the same subspace. Now, there are two subcases: First,  in each of the filled sectors, one physical and one auxiliary particle is placed, or two physical particles are placed in one sector and two auxiliary particles in the other. The first case is just two times the one particle case discussed above, and thus, it gives rise to even states. 
The second case however is different; if the two occupied sectors are labeled by $m$ and $r$, the constraint translates to 
\begin{equation}
 n_m=n_{L+1-m}=1, \quad \quad \tilde{n}_r=\tilde{n}_{L+1-r}=1 \;,
\end{equation}
which also transforms evenly. This concludes the discussion in the case of an even number of sites.\\
For odd $L$, the case distinction above stays valid (although the arguments, why in each case the state transforms evenly, change). However, due to the subtlety of the zero\hyp sector and $E_{k(\nicefrac{L+1}{2})}=0$, particles can sit in the mode $k(\nicefrac{L+1}{2})$ without fulfilling any constraint. 
Luckily, to satisfy the other sector constraints, an even number of particles has to sit in this zero\hyp sector (as the physical plus auxiliary particle number is even). Consequently, either all particles are in the zero\hyp sector, or two particles are in there, and the remaining two particles need to obey the fillings discussed in the one-particle case. So, as every single particle in the zero\hyp sector has odd symmetry (but an even number of them is in there), we only get even symmetry states.\\

Now, before this theorem is used, a neat symmetry property of the Lindbladian needs to be shown.\\

\textit{4. Theorem}: The Lindbladian preserves the inversion symmetry $\hat{\mathcal{U}}_{\mathrm{inv}}$.\\

\noindent This is immediately clear for the Hamiltonian part. For the dissipator, however, we have 
\begin{align}
    &\hat{U}_\text{inv}^\nodagger\hat{L}_j^\nodagger\hat{U}_\text{inv}^\dagger = \hat{U}_\text{inv}^\nodagger\sqrt{\kappa}(\hat{b}^\dagger_{j+1}+\hat{b}^\dagger_j)\hat{U}_\text{inv}^\dagger\hat{U}_\text{inv}^\nodagger(\hat{b}_{j+1}^\nodagger-\hat{b}_j^\nodagger)\hat{U}_\text{inv}^\dagger \nonumber\\
    &= \sqrt{\kappa}(\hat{b}^\dagger_{L-j}+\hat{b}^\dagger_{L-j+1})(\hat{b}_{L-j}^\nodagger-\hat{b}_{L-j+1}^\nodagger) = -\hat{L}_{L-j}^\nodagger\; ,
\end{align}
which shows preservation of symmetry, as in the dissipator only products of two jump operators exist, and a sum over all sites is performed.\\

Now we can bring everything together to conclude our main theorem.\\

{\textit{5. Main theorem}: Consider the Lindbladian with $k_0=0$. All left eigenmodes, which, when adiabatically switching on the dissipation, stem from the eigenspace of $\hat{\mathcal{H}}$ to eigenvalue $-\mathrm i\big(E_{k(1)}-E_{k(2)}\big)$, will have zero overlap with symmetric states.\\

\textit{Proof of 5}: Let $\hat{l}_m$ be an eigenmode coming out of the eigenspace $-\mathrm i(E_{k(1)}-E_{k(2)})$. That is, there is a $\hat{l}^{\text{appr.}}_m$ in this subspace which agrees with $\hat{l}_m$ in the case of vanishing dissipation $\kappa\to0^+$. Such an approximate mode is found by (degenerate) perturbation theory and we can therefore safely say that $\text{Tr}\big{(}\hat{l}^\dagger_m\hat{l}^{\text{appr.}}_m\big{)} \neq 0$ also for strong dissipation (see \cref{fig:perturbation:theory}). However, since the approximate mode lives in a space spanned by only antisymmetric states according to theorem 1, $\hat{l}_m$ has an antisymmetric component. By theorem 4, the Lindbladian preserves the inversion symmetry, so that the left eigenmodes can be chosen as eigenvectors of $\hat{\mathcal{U}}_\textrm{inv}$. We conclude, that $\hat{l}_k$ has to be antisymmetric and thus has vanishing overlap with symmetric states.\\

Notice, that oddly transforming states are traceless, and thus, all physical states have an evenly transforming component. 
Luckily, there are density matrices that transform evenly, making them orthogonal to eigenmodes coming out of the eigenspace, as discussed above.
%

%
% We can generalize the $U$, For this, notice that the part of the Hamiltonian representing the onsite interaction also preserves the inversion symmetry. Thus, the main theorem stays valid also in the $U\neq 0$ case.

Unfortunately, the main theorem tells nothing about the location of the eigenvalues corresponding to these left eigenmodes.
Thus, it remains to show that the slowest decaying mode actually stems from the correct eigenspace of $\hat{\mathcal H}$. Notice that we can tackle the case of finite $U$ in the same fashion, by slowly turning on $U = \mathcal O(\kappa)$. 

\begin{figure}[ht!]
    \centering
    \subfloat[\label{subfig:supmat:evals:a}]{
        \includegraphics[width=0.94\columnwidth]{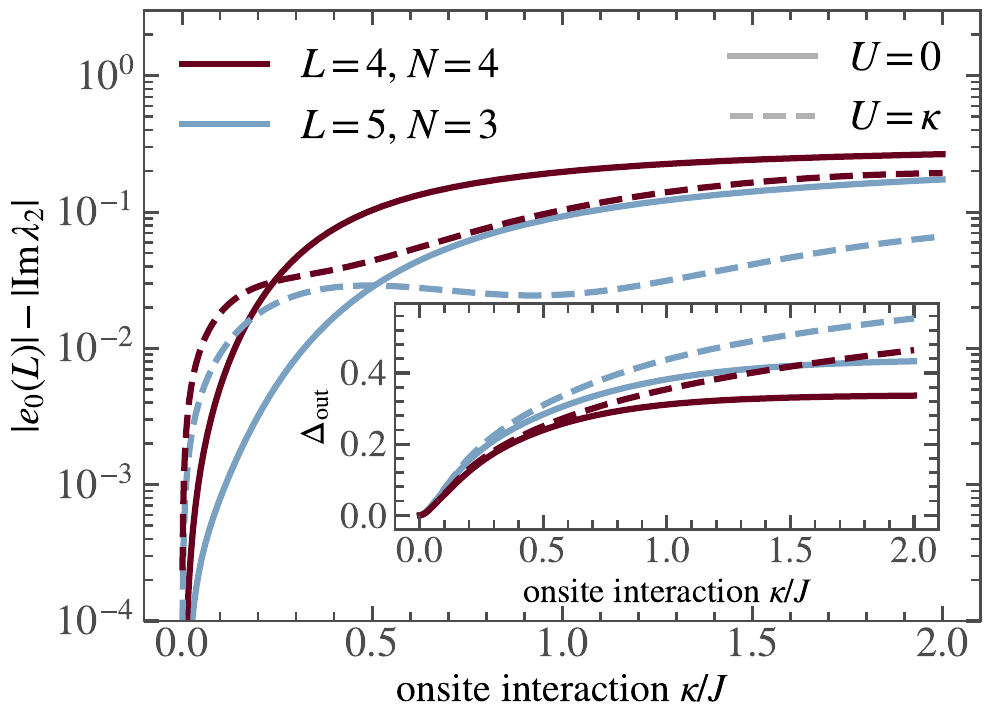}
 }
    \vspace{-1em} 
    % Panel (c)
    \subfloat[\label{subfig:supmat:overlap:b}]{
        \includegraphics[width=0.98\columnwidth]{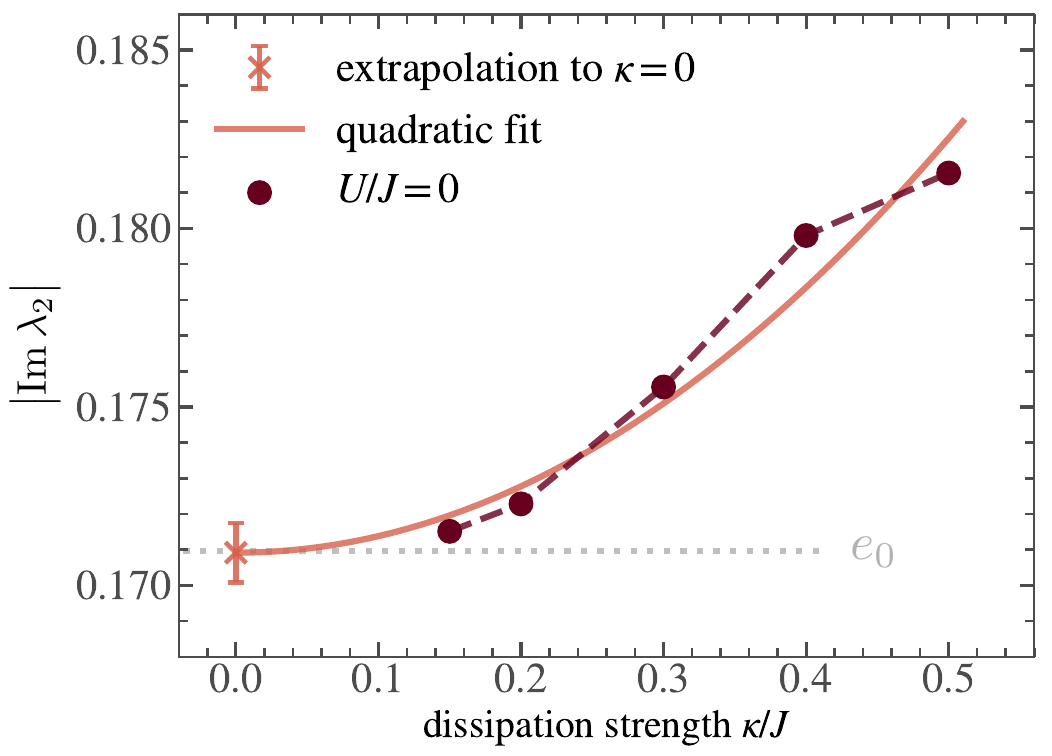}
 }
    \caption{Characteristics of the slowest decaying mode for different system parameters. Panel a): Difference of $\text{Im}\lambda_2$ to the subspace eigenvalue $e_0(L)=E_{k(2)}-E_{k(1)}$ (see \cref{eq:app:eigenvals:vectorized:H} and (\ref{eq:app:diagonalized:H})), which vanishes as $\kappa\to 0^+$. \updated{Inset: Component $\Delta_\mathrm{out}$ of the eigenmode that lies outside the eigenspace corresponding to the eigenvalue $\mathrm i \,e_0(L)$ (see \cref{eq:bec:outside:weight}). $\Delta_\mathrm{out}$ approaches 0 for $\kappa \to 0^+$, which shows that $\hat{l}_2$ stems from this eigenspace and $\hat{l}_2$ is oddly transforming. Note also that the perturbation theory is still valid when slowly increasing the onsite interaction $U$ with $\kappa$. The data has been generated with $J=1$ and using \acrfull{ED}.
    Panel b): Analysis for a system with $L=12$, $N=8$ using \gls{CLIK-MPS}~\cite{westhoff2025tensor}. We calculate $\big\lvert \mathrm{Im} \, \lambda_2 \big\lvert$ and fit it with the ansatz $a_1 + a_2\kappa^2$, which is motivated from the \gls{ED} analysis. We recover the value $a_1 = 0.1709 \pm 9\times10^{-4}$, which is in agreement with $e_0 = 0.1710$, proving the emergence of the quantum Mpemba effect in this large scale system.
    We set $U/J=0$ and chose the parameters for \gls{CLIK-MPS} such that convergence is reached.}
 }
    \label{fig:perturbation:theory}
\end{figure}
\updated{
There are two signatures that tell whether the slowest decaying mode stems from the eigenspace in question or not.
If we denote by $\{ \vecket{e_0, p} \}_p$ the eigenvectors of $\hat{\mathcal H}$ to eigenvalue $e_0$, with $p$ numbering the multiplicity, we can compute the overlap of the eigenvector $\hat l_2$ with the eigenspace of $\hat{\mathcal H}$. 
This is done by constructing the projector onto the eigenspace $\hat \Pi_{e_0} = \sum_p \vecket{e_0, p}\vecbra{e_0, p}$, and projecting the slow decaying eigenvector onto this subspace. The component orthogonal to the eigenspace is then given by} 
\begin{equation}
    \updated{\Delta_\mathrm{out}(\kappa) = 1- \sum_p \big\vert \vecbraket{e_0, p}{l_2(\kappa)}\big\vert^2 \; .}
    \label{eq:bec:outside:weight}
\end{equation}
\updated{Crucially, there exists a normalized vector $\vecket{l_2^\mathrm{appr.}} = \sum_l c_l \vecket{e_0, l}$ with weights $c_l$ such that $\Delta_\mathrm{out} = 1-\vert\vecbraket{l_2^\mathrm{appr.}}{l_2}\vert^2$. The state $\vecket{l_2^\mathrm{appr.}}$ is the best approximation in the subspace since it maximizes the overlap with $\vecket{l_2}$. 
A hard proof for the state $\vecket{l_2}$ stemming from the eigenspace in question is thus the limit behavior}
\begin{equation}
    \updated{\lim_{\kappa\to 0^+} \Delta_\mathrm{out}(\kappa) = 0 \; .}
    \label{eq:bec:limit:behavior:subspace:2}
\end{equation}
\updated{To check \cref{eq:bec:limit:behavior:subspace:2} in a large system requires to know $\vecket{l_2}$ and a full basis of the (potentially very large) subspace to the eigenvalue $e_0$.
A handier criterion is connected to the imaginary part of $\lambda_2$. If it stems from the subspace, we should recover}
\begin{equation}
    \updated{\lim_{\kappa\to0^+} \vert\mathrm{Im} \, \lambda_2\vert = E_{k(1)} - E_{k(2)} = e_0 \; ,}
    \label{eq:bec:limit:behavior:subspace:1}
\end{equation}
\updated{Indeed, we can establish equivalence between the two criteria.
As already stated, it is immediately clear that \cref{eq:bec:limit:behavior:subspace:2} implies \cref{eq:bec:limit:behavior:subspace:1}. To prove the other direction, we start by explicitly separating the $\kappa$ dependence from the Lindbladian via $\hat{\mathcal L} = \hat{\mathcal H} + \kappa\hat{\mathcal D}$. The proof is based on the observation that}
\begin{equation}
    \updated{\vecbra{l_2(\kappa)} \big( \hat{\mathcal H} - E(\kappa)\hspace{-2pt} \big)^2\vecket{l_2(\kappa)} = \mathcal{O}(\kappa) \; ,}
\end{equation}
\updated{where we introduced new notation for the Hamiltonian expectation value $E(\kappa) = \vecbra{l_2(\kappa)} \hat{\mathcal H}\vecket{l_2(\kappa)}$.
By calculating the variance in the eigenbasis of $\hat{\mathcal H}$ and using it to bound $\Delta_\mathrm{out}(\kappa)$, as well as showing the convergence of $E(\kappa) \to e_0$ as $\kappa\to0^+$ we end up with}
\begin{equation}
    \updated{\Delta_\mathrm{out}(\kappa) = \mathcal{O}(\kappa) \; ,}
\end{equation}
\updated{concluding the proof of equivalence. 
Hence, it suffices to check \cref{eq:bec:limit:behavior:subspace:1}, which directly implies the eigenmode stemming from the correct eigenspace of $\hat{\mathcal H}$.
In \cref{fig:perturbation:theory} we check the criteria \cref{eq:bec:limit:behavior:subspace:2} and (\ref{eq:bec:limit:behavior:subspace:1}).
First, in \cref{subfig:supmat:evals:a} we study the behavior \cref{eq:bec:limit:behavior:subspace:1} in systems with $L=4, N=4$ (red) and $L=5$, $N=3$ (blue). For $J=1$ and $U=0$ (solid lines) we see the difference of $\mathrm{Im} \, \lambda_2$ and $e_0$ vanish upon lowering $\kappa$. We furthermore find that the difference vanishes quadratically as $\big\lvert\mathrm{Im}\, \lambda_2 \big\lvert = e_0 + \mathcal{O}(\kappa^2)$, which will be important for the large scale analysis.
Similar behavior is found when considering a finite $U$ (dashed lines). As described before, we turn on $U$ slowly by considering $U=\kappa$. 
In the inset we check \cref{eq:bec:limit:behavior:subspace:2} for the same parameters, and find $\Delta_\mathrm{out} \to 0$, as predicted by the equivalence of the criteria \cref{eq:bec:limit:behavior:subspace:2} and (\ref{eq:bec:limit:behavior:subspace:1}).}
\updated{To check if the analytic arguments also apply at larger system sizes not accessible to \gls{ED} methods, we can employ CLIK-MPS to calculate $\lambda_2$ \cite{westhoff2025tensor}.}
\begin{figure*}[ht!]
    \centering
    \subfloat[\label{subfig:bec:phase:a}]{
        \includegraphics[width=0.302\linewidth]{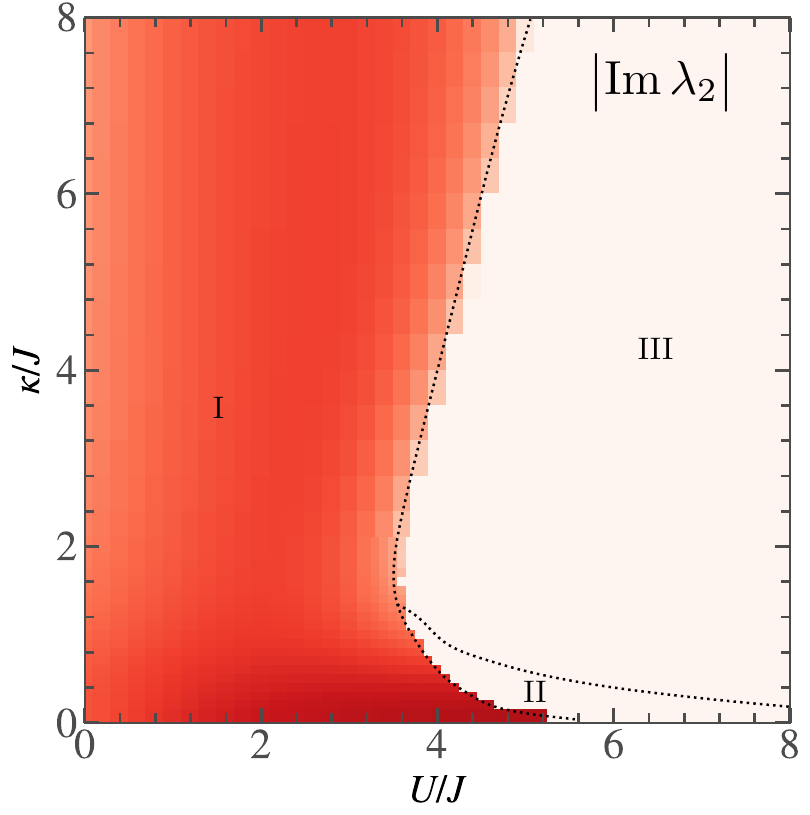}
    }
    \subfloat[\label{subfig:bec:phase:b}]{
        \includegraphics[width=0.302\linewidth]{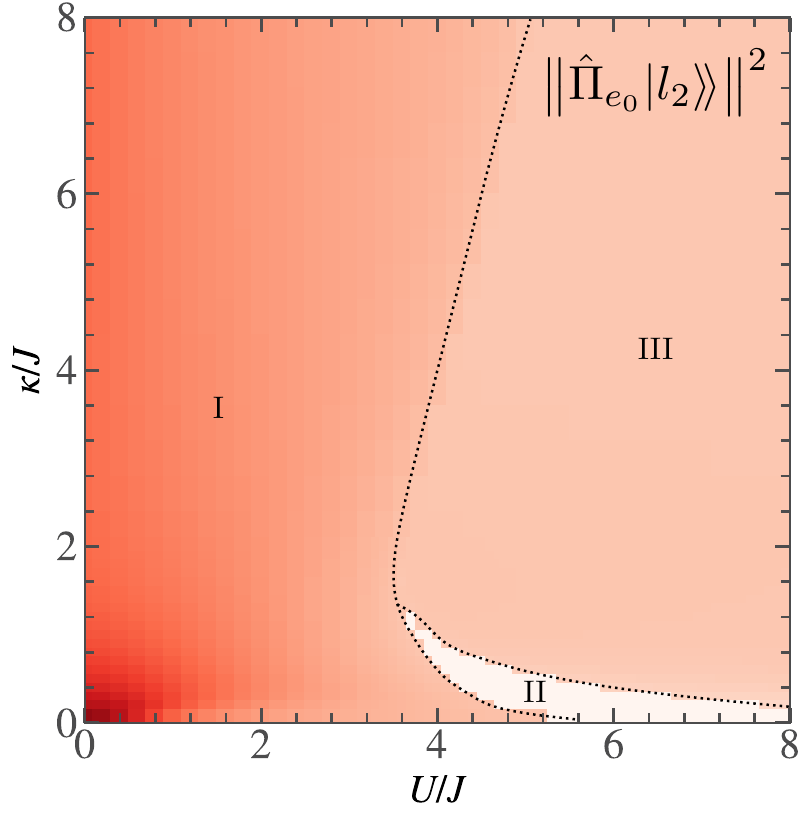}
    }
    \subfloat[\label{subfig:bec:phase:c}]{
        \includegraphics[width=0.365\linewidth]{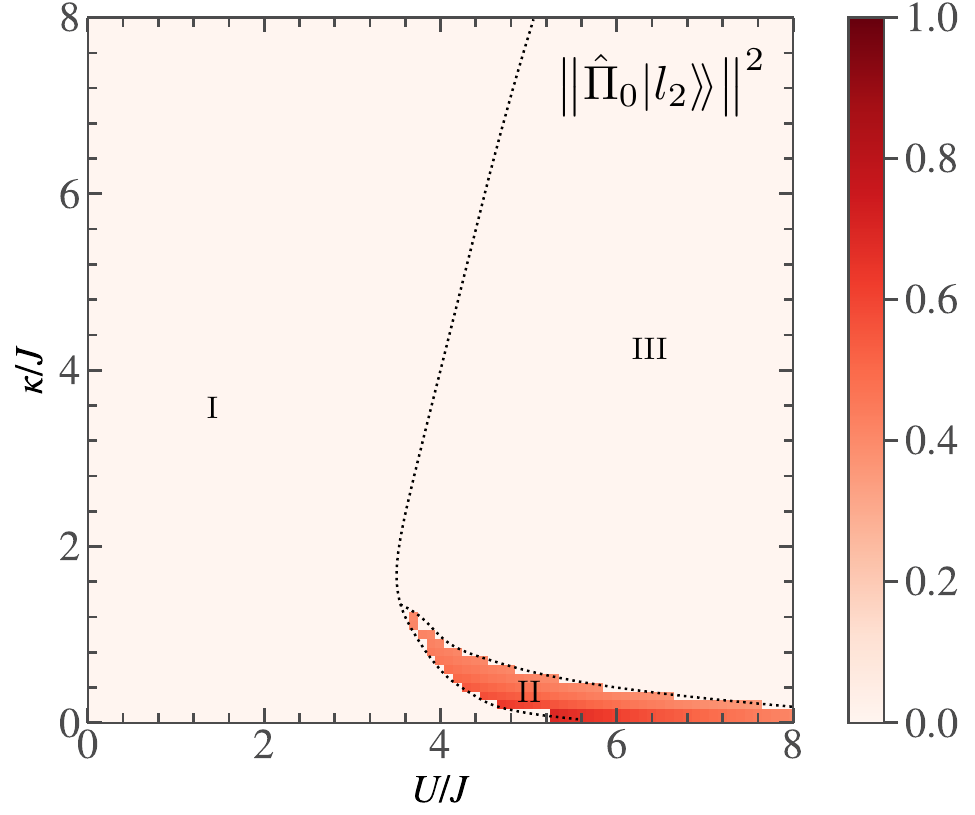}
    }
    \hfill
    \caption{%
\updated{Slowest decaying mode of the dissipative Bose Hubbard model with $L=6$ and $N=3$. 
Panel a): Modulus of $\mathrm{Im}\,\lambda_2$. 
Panel b): Overlap of the slowest decaying left eigenvector with the $e_0$\hyp eigenspace, which only contains oddly transforming states (c.f. theorem 1).
Panel c): Overlap of the slowest decaying left eigenvector with the $0$\hyp eigenspace, which only contains evenly transforming states (c.f. theorem 2).
The dotted black lines mark sudden changes in the observed quantities, and separate the regions I, II and III. All calculations were performed using \gls{ED}.}
}
    \label{fig:bec:phases}
\end{figure*}

\updated{%
In \cref{subfig:supmat:overlap:b} we show $\big\lvert\mathrm{Im}\, \lambda_2 \big\lvert$ for a system with 12 sites and 8 bosons at $U/J=0$ depending on the dissipation strength $\kappa/J$ calculated via \gls{CLIK-MPS} \cite{westhoff2025tensor}. 
Upon lowering the dissipation strength, $\big\lvert\mathrm{Im}\, \lambda_2 \big\lvert$ approaches the subspace eigenvalue $e_0 = 0.1710$. To assess if it actually satisfies \cref{eq:bec:limit:behavior:subspace:1}, we can use the insights from the \gls{ED} analysis and extrapolate the data to $\kappa=0$ using $\big\lvert\mathrm{Im}\, \lambda_2 \big\lvert = a_1 + a_2\kappa^2$. This reveals $a_1 = 0.1709 \pm 9\times10^{-4}$, which is in agreement with $e_0$, and proves that the analytic results to explain the emergence of the Mpemba effect also apply in large scale systems.}
Note that the analytic arguments can be generalized to higher dimensions. For instance, in 2D, it can be shown that the Lindbladian again features inversion symmetries of the two spatial coordinates. 
The main difference to the 1D case lies in the spectrum of the Hamiltonian, whose degeneracy substantially increases with the dimension. This makes the proof of the main theorem more involved and will be the subject of future studies.
Nonetheless, \gls{ED} results for small-scale systems indicate, that the \gls{SL} state which we considered in the main text, converges to the \gls{BEC} exponentially faster than typical random states also in 2D.

\updated{\section{Existence of symmetry\hyp based Mpemba speedups in different parameter regimes}}
\updated{%
To assess if the Mpemba effect emerges in all parameter regimes, we perform \gls{ED} calculations in \cref{fig:bec:phases} for various $\kappa/J$ and $U/J$ for a system of $L=6$ and $N=3$ particles. 
In \cref{subfig:bec:phase:a} we show the modulus of the imaginary part of the slowest decaying mode, in \cref{subfig:bec:phase:b} the overlap of $\vecket{l_2}$ with the $e_0$\hyp eigenspace of $\hat{\mathcal H}$ and in \cref{subfig:bec:phase:c} the overlap of $\vecket{l_2}$ with the $0$\hyp eigenspace of $\hat{\mathcal H}$.
As predicted by our perturbative analysis in \cref{sec:app:fast:states}, for small $U\ll \max(J, \kappa)$, $\big\vert \mathrm{Im} \, \lambda_2\big\vert$ is strictly positive, which leads to a finite overlap with the $e_0$\hyp eigenspace and thus an odd transformation behavior under $\hat{\mathcal U}_\mathrm{inv}$.
Increasing $U/J$, $\big\vert \mathrm{Im} \, \lambda_2\big\vert$ abruptly jumps to zero, and we expect $\hat l_2$ to transform evenly. This indeed happens for small $\kappa/J$, which is seen in the overlap with the $0$\hyp eigenspace in \cref{subfig:bec:phase:c}, where the overlap jumps from zero to a finite value.
Interestingly though, at higher $U/J$ or $\kappa/J$ the overlap with the $0$\hyp eigenspace jumps back to zero, although $\big\vert \mathrm{Im} \, \lambda_2\big\vert$ is still zero. In this region, the overlap with the $e_0$\hyp eigenspace is again finite, indicating an odd transformation behavior and the emergence of the Mpemba effect. 
Summarizing, there are the following three spectral regions, which are all differnetiated through jumps in an observable.}
\vspace{0.2cm}
\begin{itemize}
    \item \updated{Region I: \textit{Finite} $\big\vert \mathrm{Im} \, \lambda_2\big\vert$, $\vecket{l_2}$ stemming out of the $e_0$\hyp eigenspace, thus \textit{odd} transformation behavior. Admitting Mpemba speedups and perturbative description.}
    \item \updated{Region II: \textit{Zero} $\big\vert \mathrm{Im} \, \lambda_2\big\vert$, $\vecket{l_2}$ stemming out of the $0$\hyp eigenspace, thus \textit{even} transformation behavior. Admitting no Mpemba speedups, but a perturbative description is possible.}
    \item \updated{Region III: \textit{Zero} $\big\vert \mathrm{Im} \, \lambda_2\big\vert$, but $\vecket{l_2}$ does transform \textit{oddly}. Admitting Mpemba speedups, but a perturbative description is not possible.}
\end{itemize}
\updated{Luckily, in regions I \& III the quantum Mpemba effect due to the odd transformation behavior is present, and these phases make out the biggest part of the parameter space. 
Nevertheless, only for region I the perturbative analysis provided in \cref{sec:app:fast:states} is valid. 
The emergence of the Mpemba effect in region III will be subject to future research. We believe that it may be also described using perturbation theory, but starting from the case $\kappa=0, \, U=\infty$.
}
\updated{It is important to verify that region II does not expand with increasing lattice size, so that the Mpemba effect is stable. We again employ \gls{CLIK-MPS} \cite{westhoff2025tensor}, which makes it possible to calculate low-lying eigenvalues and overlaps with the corresponding eigenvectors.
Since $\lambda_2$ can be approximated within \gls{CLIK-MPS}, \cref{subfig:bec:phase:a} can be recreated for larger systems. Unfortunately, the projections needed for \ref{subfig:bec:phase:b} and \ref{subfig:bec:phase:c} are prohibitly difficult to calculate for large systems and we directly consider the transformation behavior of $\hat l_2$ under $\hat{\mathcal U}_\mathrm{inv}$. As already established, it is even in II but odd in I \& III. Notice that it suffices to examine the transformation behavior of the right eigenvector $\vecket{r_2}$, since $\vecbraket{r_2}{l_2}\neq 0$.
We generate a random product state $\vecket{\psi}$ and calculate the overlap of $\vecket{r_2}$ with its projection onto the evenly and oddly transforming subspace, denoted by $\psi_\mathrm{even/odd}$. The overlap ratio $R_\mathrm{ov}(\psi)$ is then defined by}
\begin{equation}
    \updated{R_\mathrm{ov}(\psi) = \frac{\big\lvert\vecbraket{\psi_\mathrm{even}}{r_2}\big\lvert^2}{\big\lvert\vecbraket{\psi_\mathrm{odd}}{r_2}\big\lvert^2} \, .}
    \label{eqn:bec:overlap:ratio}
\end{equation}
\updated{We take the mean overlap ratio $R_\mathrm{ov}$, evaluated over many random states, for the calculations provided here we found 50 realizations to be sufficient.
If the right eigenmode transforms oddly, we expect $R_\mathrm{ov}=0$ while for even transformation behavior $R_\mathrm{ov}=\infty$. We thus distinguish between the casew $R_\mathrm{ov}\ll1$ and $R_\mathrm{ov}\gg1$, since CLIK-MPS only yields approximate eigenmodes.}
\begin{figure}[t]
    \centering
    \subfloat[\label{subfig:bec:clik:phase:a}]{
    \includegraphics[width=0.953\columnwidth]{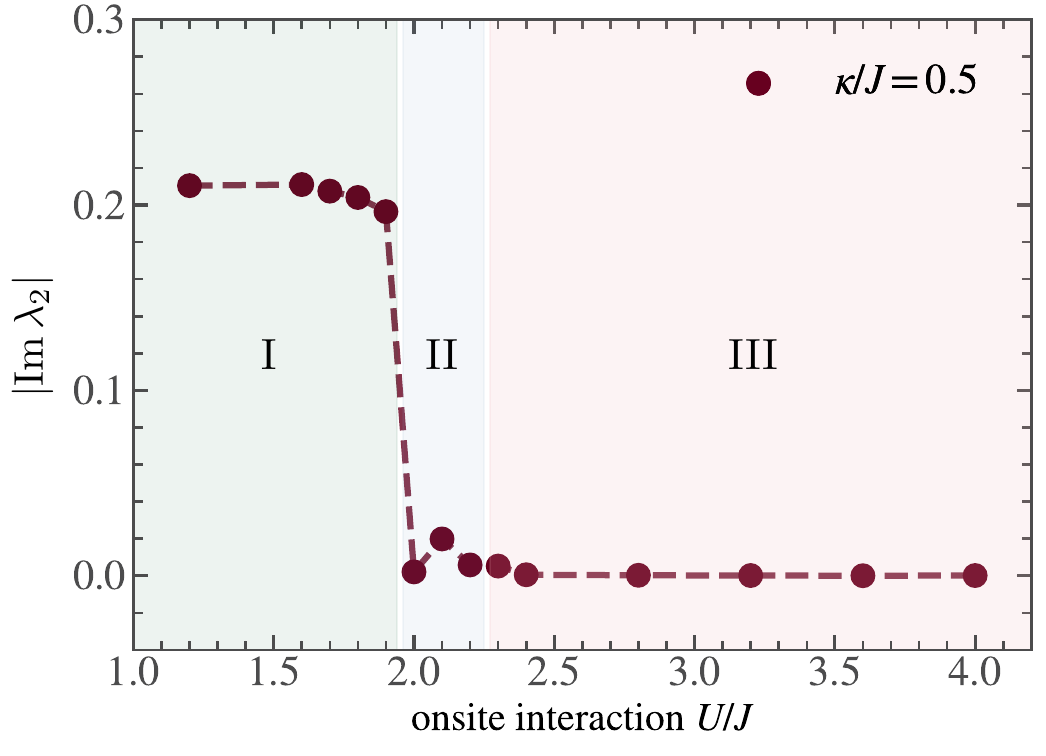}
    }
    \vspace{-1em}
    \subfloat[\label{subfig:bec:clik:phase:b}]{
        \includegraphics[width=0.983\columnwidth]{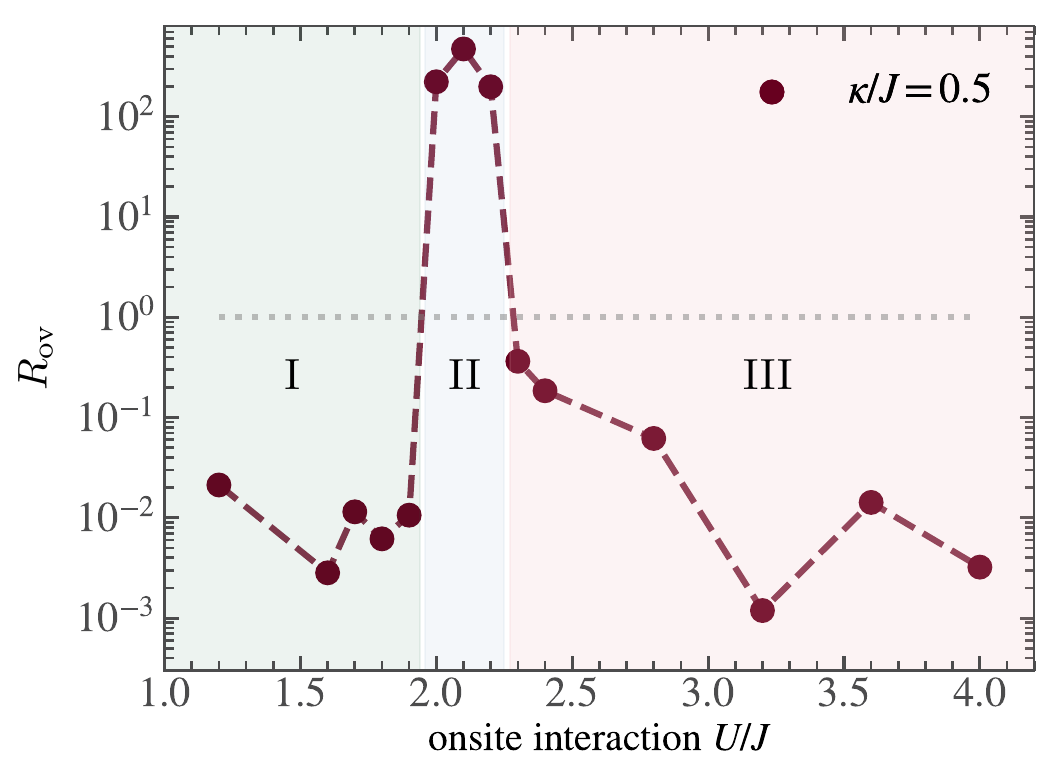}
    }
    \hfill
    \caption{%
\updated{Different parameter regimes at $\kappa/J=0.5$ for a large-scale system with 16 sites and 8 bosons. 
Panel a): Modulus of $\lambda_2$'s imaginary part. 
Panel b): Overlap ratio $R_\mathrm{ov}$ (see \cref{eqn:bec:overlap:ratio}) averaged over 50 random product states.
The shaded areas highlight region I (green), II (blue) and III (red). All data was obtained using \gls{CLIK-MPS} with a sampling rate $\delta t = 0.4$ and maximal time $T=40$ with complex angle $\alpha = 0.02$ in addition to the real time evolution. The local dimension was set to $d=9$.}
}
    \label{fig:bec:clik:phases}
\end{figure}
\updated{In \cref{fig:bec:clik:phases} we consider a system of 16 sites and 8 bosons, with $\kappa/J=0.5$, which should place us in a regime where all three phases are visible when varying $U/J$, according to \cref{fig:bec:clik:phases}. 
If we turn on the onsite interaction carefully, we find a finite $\mathrm{Im}\, \lambda_2$ in \cref{subfig:bec:clik:phase:a} combined with an odd transformation behavior in \cref{subfig:bec:clik:phase:b} up until $U/J =1.95\pm 0.05$. We thus identify this as region I. 
At $U/J =1.95\pm 0.05$ the behavior abruptly changes, $\mathrm{Im}\, \lambda_2$ drops to zero and $R_\mathrm{ov}\gg1$, indicating the transition to region II. 
The mean overlap ratio again drops to $R_\mathrm{ov}\ll1$ at $U/J=2.25\pm0.05$, while $\mathrm{Im}\, \lambda_2$ does not change, showing that the slowest decaying mode again transforms oddly, and indicating the onset of region III. 
Notice how the behavior does not qualitatively differ from the small system discussed in \cref{fig:bec:phases}, and most importantly, the size of region II at fixed $\kappa/J=0.5$ has not increased when comparing to the $L=6$ data. This suggests that a symmetry\hyp based Mpemba speedups will be found for most values of the parameters $(U,\kappa)$ also for very large system sizes.}
\section{Unitary equivalence between Lindbladians with different momenta}
\label{sec:app:unitary:equivalence}
In \cref{sec:app:fast:states}, we only considered Lindbladians with zero characteristic momentum $k_0=0$. 
Instead of simply generalizing the arguments given above to arbitrary $k_0$ (which is possible), here we show that an even stronger connection exists between the Lindbladians with different $k_0$. This is done by introducing unitaries, transforming Lindbladians with different parameters into one another.
As a consequence, they share the same spectrum and their eigenmodes are connected.

Consider the unitary $\hat{U}_{k_0}^\nodagger$, acting on creation operators as
\begin{equation}
    \hat{U}_{k_0}^\nodagger \hat{b}_j^\dagger\hat{U}_{k_0}^\dagger = \mathrm e^{\mathrm ik_0j}\hat{b}_j^\dagger \, .
\end{equation}
This unitary has the physical interpretation of shifting the momentum by $k_0$.
To identify the transformation behavior of the Lindbladian, we study the Hamiltonian and jump operators separately.
The Hamiltonian transforms as 
\begin{align}
    \hat{U}_{k_0}^\nodagger\hat{H}_{k=0}(J, U)\hat{U}_{k_0}^\dagger = \hat{H}^\nodagger_{k=k_0}\big(\sigma_{k_0}J, U\big)\; ,
    \label{eq:app:trafo:for:ham}
\end{align}
and the jump operators obey
\begin{align}
    \hat{U}_{k_0}^\nodagger\hat{L}_{j}^{k=0}\hat{U}_{k_0}^\dagger = \hat{L}_j^{k=k_0} \, .
\end{align}
Thus, the Lindbladians at different momenta are unitarily connected through the unitary $\hat{\mathcal{U}}_{k_0}^\nodagger= \hat{U}_{k_0}^\nodagger \otimes \hat{U}_{k_0}^{*}$, which is the vectorized version $\hat{U}_{k_0}$. 
As a consequence, the Lindbladian spectra are independent of the targeted momentum $k_0$.
The eigenmodes are also related as
\begin{equation}
    \vecket{r_j^{k=k_0}} = \hat{\mathcal{U}}_{k_0}^\nodagger \vecket{r_j^{k=0}}, \; \vecbra{l_j^{k=k_0}} = \vecbra{l_j^{k=0}} \hat{\mathcal{U}}_{k_0}^\dagger \, .
\end{equation}
The one\hyp body density matrix $\bm\gamma_{ij} = \langle\hat{b}^\dagger_j\hat{b}_i\rangle$ is a valuable tool for quantitatively assessing the properties of the steady state. In the case of a finite\hyp momentum BEC, it gives information about the condensate density and the momentum through its spectral decomposition. 
Due to the unitary relation between the right eigenmodes, the properties of the steady state's one body density matrix are also related. By inserting the unitary relation, using the cyclicity of the trace and the definition of $\hat{U}_{k_0}$, one finds
\begin{equation}
    \bm{\gamma}_{lm}^{k_0} = \text{Tr}\big{(} \hat{r}_{1}^{k_0} \hat{b}_m^\dagger \hat{b}_l^\nodagger\big{)} = e^{ik_0(m-l)} \text{Tr}\big( \hat{r}_{1}^{0} \hat{b}_m^\dagger \hat{b}_l^\nodagger\big) \, .
\end{equation}
Fourier\hyp transforming $\bm\gamma$ to momentum space gives
\begin{equation}
    \bm{\gamma}^{k_0}_{k, k'} = \bm{\gamma}^0_{k+k_0, k'+k_0} \, ,
\end{equation}
which shows a unitary connection between $\bm\gamma^{k_0}$ and $\bm\gamma^0$. This relation is also consistent with the physical interpretation of the unitary transformation, shifting all momenta by $k_0$. Thus, the eigenvalues are the same, and the eigenmodes are connected by an index shift of $k_0$. Consequently, the condensate density only depends on wether $|k_0|<\pi/2$ or $|k_0|\geq \pi/2$, as well as the system parameters $J$, $U$ and $\kappa$.

Furthermore, the unitary affects the symmetry properties of $\hat{l}_2^{k_0}$, which can be deduced from the inversion symmetry present in the $k_0=0$ case and the unitary transformation. This gives rise to a symmetry of the Hamiltonian, which depends on the targeted momentum $k_0$,
\begin{align}
    \hat{U}_{\text{inv}}^{k_0} = \hat{U}^\nodagger_{k_0}\hat{U}^\nodagger_{\text{inv}}\hat{U}_{k_0}^\dagger \, .
\end{align}
Its action on creation operators is given by
\begin{equation}
    \hat{U}_{\text{inv}}^{k_0}\hat{b}_j^\dagger\big{(}\hat{U}_{\text{inv}}^{k_0}\big{)}^\dagger = \mathrm e^{\mathrm ik_0(L+1)}\mathrm e^{-2\mathrm ik_0 j}\hat{b}_{L+1-j}^\dagger \, .
\end{equation}
The symmetric product states introduced before are again eigenstates of this symmetry and transform evenly as before.
Also, $\hat{l}_2^{k_0}$ again transforms oddly, which implies that symmetric states will equilibrate exponentially faster than random states for arbitrary momenta $k_0$.\\

The Lindbladians for $|k_0|<\pi/2$ and $|k_0|\geq\pi/2$  are also connected unitarily; however, the hopping amplitude changes sign. 
To show this, we need to introduce a further unitary, which moves the changed sign in \cref{eq:app:trafo:for:ham} to the onsite interaction $U$.
For this, we can restrict our analysis to the case of $k_0=0$.
This time, it is beneficial to work directly on the vectorized Hilbert space.
To avoid confusion, we will denote creation (annihilation) operators on the physical sublattice by $\hat b^\dagger$ ($\hat b$), and on the auxiliary sublattice by $\hat a^\dagger$ ($\hat a$).
We define the unitary $\hat{\mathcal{S}}$
\begin{equation}
    \hat{\mathcal{S}} \hat{a}_j \,\hat{\mathcal{S}}^\dagger = \hat b_j\; , \quad \hat{\mathcal{S}} \hat{b}_j \,\hat{\mathcal{S}}^\dagger = \hat a_j \; .
\end{equation}
Denoting by $\hat{\mathcal{H}}_0(J, U)$ (we write the parameter dependence explicitly here, since it plays a major role) and $\hat{\mathcal{D}}_0$ the Hamiltonian and dissipative part of the Lindbladian $\hat{\mathcal{L}}_0$, respectively, we find the transformation behavior
\begin{equation}
    \hat{\mathcal{S}}\hat{\mathcal{H}}_0(J, U)\hat{\mathcal{S}}^\dagger =\hat{\mathcal{H}}_0( -J, - U) \; ,
\end{equation}
where we used the relation $\hat H_0^T(J, U) = \hat H_0(J, U)$.
Notice that both $J$ and $U$ have changed sign.
Next, $\hat L_j^0 = \big[L_j^0\big]^*$, where $*$ denotes a complex conjugation, can be utilized to show
\begin{equation}
    \hat{\mathcal{S}}\hat{\mathcal{D}}_0\hat{\mathcal{S}}^\dagger =\hat{\mathcal{D}}_0 \; .
\end{equation}
This finally yields unitary equivalence between the Lindbladians 
\begin{equation}
    \hat{\mathcal{S}}\big(\hat{\mathcal{H}}_0(J, U)+\hat{\mathcal{D}}_0\big)\hat{\mathcal{S}}^\dagger =\big(\hat{\mathcal{H}}_0(-J, -U)+\hat{\mathcal{D}}_0\big) \;.
\end{equation}
Additionally, the pure product states $\vecket{\rho} \sim \prod_j \big[\hat a^\dagger_j\big]^{N_j}\big[\hat b^\dagger_j\big]^{N_j} \vecket{\text{vac}}$ are eigenstates of this unitary.
Therefore, the convergence to the respective steady state is the same for both Lindbladians.
If we now also use the unitary connection before, we find that
\begin{equation}
    \hat{\mathcal{U}}_{k_0}^\nodagger\hat{\mathcal{S}}\big(\hat{\mathcal{H}}_{0}(J, -U)+\hat{\mathcal{D}}_{0}\big)\hat{\mathcal{S}}^\dagger\hat{\mathcal{U}}_{k_0}^\dagger =\hat{\mathcal{H}}_{k_0}(J, U) + \hat{\mathcal{D}}_{k_0} \; , 
\end{equation}
and thus, the Lindbladian at positive $U>0$ and $|k_0|\geq\pi/2$ is unitarily connected to the ones with $U<0$ and $|k_0|<\pi/2$.
Moreover, $k_0 =\pi$ is especially interesting, since the Hamiltonian satisfies $\hat H_{k_0 =0} = \hat H_{k_0=\pi}$ and only the jump operators change.
Crucially, they are also real without phase modulations, and the experimental preparation of a \gls{BEC} with momentum $k_0=0$ does not require any lattice shaking.
Additionally, the uniqueness of the steady state in the zero\hyp momentum case \cite{Diehl2008} carries over to the general $k_0$ case, since the steady states are unitarily connected to one another. 
\section{Mean\hyp field theory and properties of the steady state}
In contrast to adiabatic state preparation protocols, which directly prepare the ground state of some Hamiltonian without dissipative cooling, the dissipation strength $\kappa$ serves as an additional parameter controlling the preparation accuracy.
In the following, we discuss the impact of $\kappa$ on the steady state analytically.
Let us define $n=\nicefrac{N}{L}$ and $n_0=\nicefrac{N_0}{L}$, where $N_0$ is the population of the zero\hyp momentum mode, and we use the Fourier\hyp transformed creation operators $\hat b_k^\dagger = \nicefrac{1}{\sqrt{L}}\sum_j\mathrm e^{\mathrm i kj}\hat b_j^\dagger$.
To simplify the Lindbladian specified by \cref{eq:app:bose:hubbard:ham,eq:app:jump:ops} (we again work with $k_0=0$), we assume that $\nicefrac{n-n_0}{n}=\nicefrac{N-N_0}{N} \ll 1$.
Then, we can make use of the Bogoliubov approximation $\hat b_0 = \sqrt{Ln_0}\approx\sqrt{Ln}$.
Only keeping terms in the jump operators up to order $\sqrt{n}$, discarding boundary effects and observing that the summand with momentum $k=0$ is zero yields $\hat L_k^\nodagger = 2\sqrt{n}(\mathrm e^{\mathrm ik}-1)\hat b^\nodagger_k + \mathcal{O}(1)$.
It is convenient to rotate the bosonic operators according to
\begin{equation}
    \begin{pmatrix}
        \hat c_{k}^\nodagger \\
        \hat c_{-k}^\nodagger
    \end{pmatrix}
 = \frac{1}{\sqrt 2}
    \begin{pmatrix}
        1 & 1 \\
        -1 & 1
    \end{pmatrix}
    \begin{pmatrix}
        \hat b_{k}^\nodagger \\
        \hat b_{-k}^\nodagger
    \end{pmatrix} \; ,
\end{equation}
to eliminate some couplings.
Then, keeping only terms to order $n$ in the Hamiltonian, we find
\begin{equation}
    \hat H = \sum_{k\neq 0} \bigg{\{} (\epsilon_k+Un)\hat c_k^\dagger\hat c_k^\nodagger + \frac{Un}{2} \Big{(} \big[\hat c_k^\nodagger \big]^2 + \big[\hat c_k^\dagger\big]^2 \Big{)}\bigg{\}} \, ,
\end{equation}
where we defined the non\hyp interacting one\hyp particle energies $\epsilon_k = 4J\sin^2\big(\nicefrac{k}{2}\big)$. The jump operators read
\begin{equation}
    \hat L_k= \sqrt{\kappa_k} \, \hat c_k \; , \quad \kappa_k = 16n\kappa\sin^2\big( \nicefrac{k}{2}\big) \,.
\end{equation}
Accordingly, the master equation decouples and it can be solved for each $k$ separately \cite{Diehl2008}.
The steady state is given by the mixed state 
\begin{equation}
    \hat\rho_{\mathrm{ss}} = Z^{-1} \prod_{k\neq0}\mathrm e^{-\beta_k\hat a_k^\dagger \hat a_k^\nodagger} \;,
    \label{app:eq:ss}
\end{equation}
with rotated bosonic operators $\hat a_k^\nodagger = \mathrm e^{-\mathrm i\phi_k}\cosh\theta_k \, \hat c_k^\nodagger + \mathrm e^{\mathrm i\phi_k}\sinh \theta_k \, \hat c_k^\dagger$.
The parameters are given by 
\begin{equation}
    \cosh^2\big(2\theta_k\big) = \coth^2\big(\nicefrac{\beta_k}{2}\big) = 1+\frac{\big(Un\big)^2}{(1+a^2)E_k^2} \; ,
    \label{app:eq:final}
\end{equation}
with the rescaled Bogoliubov energy $E_k = \sqrt{\epsilon_k^2+2Un\epsilon_k/(1+a^2)}$, where we defined the dimensionless constant $a=2\kappa n/J$ and the phase reads
\begin{equation}
    \cot(2\phi_k) = 2(\epsilon_k+Un)/\kappa_k \;. 
    \label{eq:app:phase:bogoliubov}
\end{equation}
%
%In principle it is not clear if these equations have solutions.
%
Note that the equation for the phase \cref{eq:app:phase:bogoliubov} always has a solution, while \cref{app:eq:final} has a solution whenever $U>0$.
Care needs to be taken in case of $U<0$, for which the constraint for existence of $\theta_k$ reads
\begin{equation}
    \sin^2\big(\nicefrac{k}{2}\big) > \frac{|U|Jn}{J^2 + \big(2n\kappa\big)^2} \; , \quad \forall \, k>\nicefrac{2\pi}{L}\; .
\end{equation}
This is fulfilled in particular in the two limits $\kappa \to \infty$ and $N \to \infty$ at constant $L$.

The steady-state \cref{app:eq:ss} is reminiscent of a thermal state.
In the limit $(Un)^2/((1+a^2)E_k^2)\gg 0$ (reformulated to $k\ll \sqrt{UJ}/\kappa$), the expression simplifies to 
\begin{equation}
    \beta_k = \frac{E_k}{T_\text{eff}} \; , \quad \quad T_\text{eff} = \frac{|U|Jn}{2\sqrt{J^2+\big( 2n\kappa\big)^2}} \, .
    \label{eq:app:effective:temp}
\end{equation}
Consequently, in this limit, the state is identical to the thermal state of an effective Bogoliubov Hamiltonian $\hat H_\text{eff}$ at effective temperature $T_\text{eff}$.
Here, $\hat H_{\text{eff}}$ is the Bose\hyp Hubbard Hamiltonian \cref{eq:app:bose:hubbard:ham} with renormalized interaction strength $U_\text{eff} = U/(1+a^2) < U$.
This approximation works particularly well for long wavelengths $k\to 0$.

We are now equipped to study the behavior of the condensate depletion.
In terms of the system parameters, it can be expressed as
\begin{equation}
    \delta = \frac{1}{N} \sum_{k\neq 0} N_k^\kappa =  \frac{1}{2N} \sum_{k} \frac{(Un)^2}{(1+a^2)E_k^2} \, .
    \label{eq:app:dissipative:depletion}
\end{equation}
Crucially, it scales as
\begin{equation}
    \label{S29}
    \delta = \mathcal{O}\big((U/\kappa)^2\;1/N\big) \; \; \; \mathrm{for} \; \; \; \kappa/J \gg \frac{L \sqrt{U}}{2\pi\sqrt{2nJ}} \;,
\end{equation}
which includes the two limits $N\to \infty$ and $\kappa \to \infty$.
This explicitly shows that a strong dissipation counters interaction effects and can drastically increase the fidelity of the prepared BEC.

Note that \cref{eq:app:effective:temp} holds in the limit $\kappa<U$. The limit $\kappa\gg U$ can be treated similarly and is relevant for comparing dissipative systems to isolated ones.
In the isolated case, the bosonic Bogoliubov operators are given by $\hat d_k = \cosh\gamma_k \, \hat c_k^\nodagger + \sinh \gamma_k \, \hat c_k^\dagger$. Here, $\gamma_k$ is related to the system parameters as
\begin{equation}
    \cosh^2(2\gamma_k) = 1 + \frac{(U_\mathrm{eff}\,n)^2}{E_k^2} \; ,
    \label{eq:isolated:bog}
\end{equation}
where $E_k = \sqrt{\epsilon_k^2+2Un\epsilon_k}$.
Then, the zero\hyp temperature equilibrium state is the groundstate of the Hamiltonian, which is exactly the vacuum of Bogoliubov quasiparticles, $\ket{\mathrm{vac}(\hat d)}$. 
The particle number can be calculated as
\begin{equation}
    N_k^{T=0} = \bra{\mathrm{vac}(\hat d)} \hat c_k ^\dagger \hat c_k ^\nodagger \ket{\mathrm{vac}(\hat d)} = \sinh^2(\gamma_k) \; .
\end{equation}
Inserting relation \cref{eq:isolated:bog} yields
\begin{equation}
    N_k^{T=0} = \frac{1}{2} \bigg( \sqrt{1+ \frac{(U_\mathrm{eff}\,n)^2}{E_k^2}} -1 \bigg) = \frac{1}{4} \frac{(U_\mathrm{eff}\,n)^2}{E_k^2} \; ,
\end{equation}
where we used the assumption $U_\mathrm{eff}\ll J$ in the last equality, as well as $L=\mathrm{const}$. 
Notice that this equation is reminiscent of \cref{eq:app:dissipative:depletion}. In the limit of low interaction $U_\mathrm{eff}\ll J/(NL)$, we can neglect some terms in the denominator and finally obtain for the isolated system at $T=0$
\begin{equation}
    N_k^{T=0} = \frac{(U_\mathrm{eff} \, n)^2}{4J^2} \frac{1}{16\sin^4(k/2)} + \mathcal{O}\big((U_\mathrm{eff} \, n/\epsilon_k)^3 \big) \:.
\end{equation}
In the dissipative case, we can approximate the particle number $N_k^\kappa$ similarly for each $k$ in the limit $\kappa\gg U$ as
\begin{equation}
    N_k^\kappa = \frac{(Un)^2}{2(J^2+4n^2\kappa^2)} \frac{1}{16\sin^4(k/2)} \; .
\end{equation}
We are interested in the effective onsite interaction $U_\mathrm{eff}$ an isolated system at zero temperature needs to have in order to generate the same depletion as a dissipatively driven system with dissipation $\kappa$. For this, we set $\delta_{T=0}(U_\mathrm{eff}) = 1/N \, \sum_k N_k^{T=0} =1/N \, \sum_k N_k^\kappa = \delta_\kappa(U)$ and rearrange to find
\begin{equation}
    U_\mathrm{eff} = \sqrt{2} \frac{U}{\sqrt{1+a^2}} \; .
\end{equation}
$U_\mathrm{eff}$ can be interpreted as the effective onsite interaction of a closed system at zero temperature with the same depletion as a dissipatively driven system with dissipation $\kappa$. 
Interestingly, the above calculation shows that even the correlation matrices of both states are the same, since the equality holds component\hyp wise, $N_k^{T=0} = N_k^\kappa$. We have
\begin{equation}
    \textrm{Tr} \big(\hat c_k^\dagger \hat c_{k'}^\nodagger \hat \rho_\mathrm{ss} \big) = \bra{\mathrm{vac}(\hat d)} \hat c_k^\dagger \hat c_{k'}^\nodagger \ket{\mathrm{vac}(\hat d)} \; .
\end{equation}
This shows that both steady states feature similar correlations.

Lastly, we are interested in the behavior of the off\hyp diagonal elements of the one\hyp body density matrix $\bm\gamma_{lm} = \langle\hat b_m^\dagger\hat b_l^\nodagger\rangle$ for the largest length scale $|l-m|=L$.
Interestingly, this is directly connected to the depletion, and we have 
\begin{equation}
    \bm{\gamma}_{lm} = n_0 + \frac{1}{L}\sum_{k\neq 0} \frac{(Un)^2}{2(1+a^2)E_k^2}\mathrm e^{\mathrm i(l-m)k} \; .
\end{equation}
Through a naive bound on the sum, we get
\begin{equation}
    \bm{\gamma}_{lm} > n(1-2\delta) = n-\mathcal O \big((U/\kappa)^2\big) \; .
    \label{eq:app:long:range:order}
\end{equation}
This can be understood as a long\hyp range order in the limits $\kappa\to \infty$ or equivalently $N \to \infty$ at constant $L$. As we show in \cref{fig:app:long:range:order}, this behavior is also found when simulating the dynamics with the full Lindbladian.

Notice that this analysis carries over to the finite $k_0$\hyp case, because of the unitary equivalences discussed in \cref{sec:app:unitary:equivalence} and the validity of the approximation for general $U\neq 0$ (at least in the parameter regimes mentioned above).
\begin{figure}
    \centering
    \subfloat[]{
    \includegraphics[width=\columnwidth]{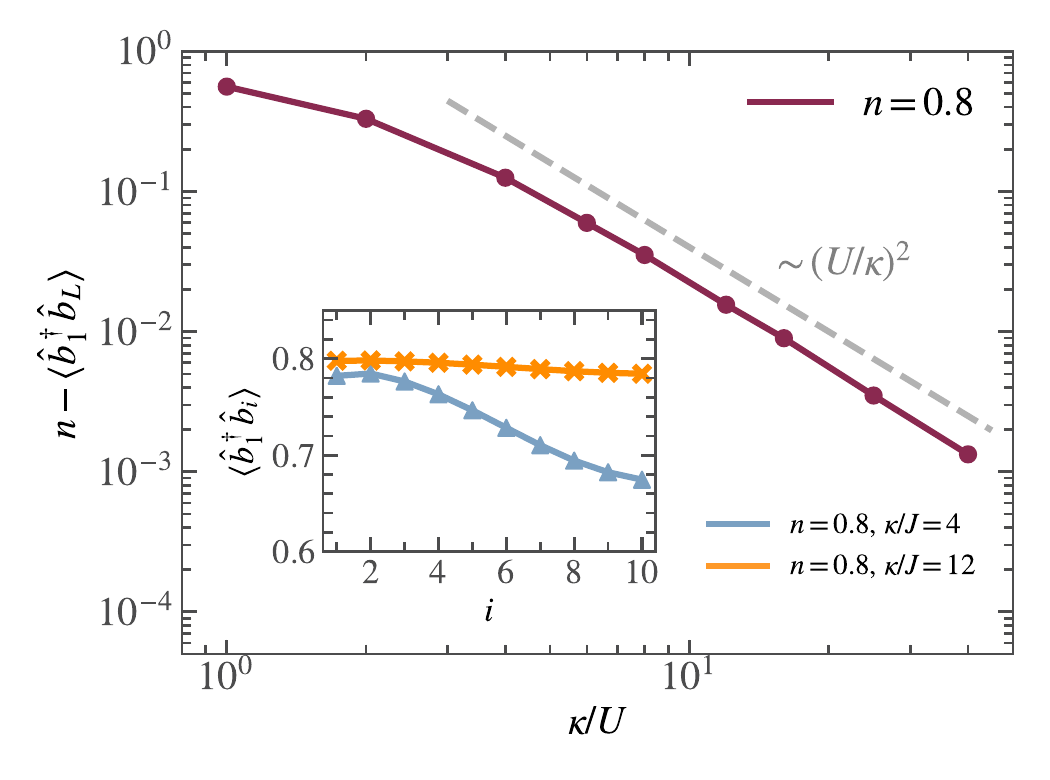}
    }
    \vspace{-0.7cm}
    \subfloat[\label{subfig:depletion:T:dependence}]{
    \includegraphics[width=0.95\columnwidth]{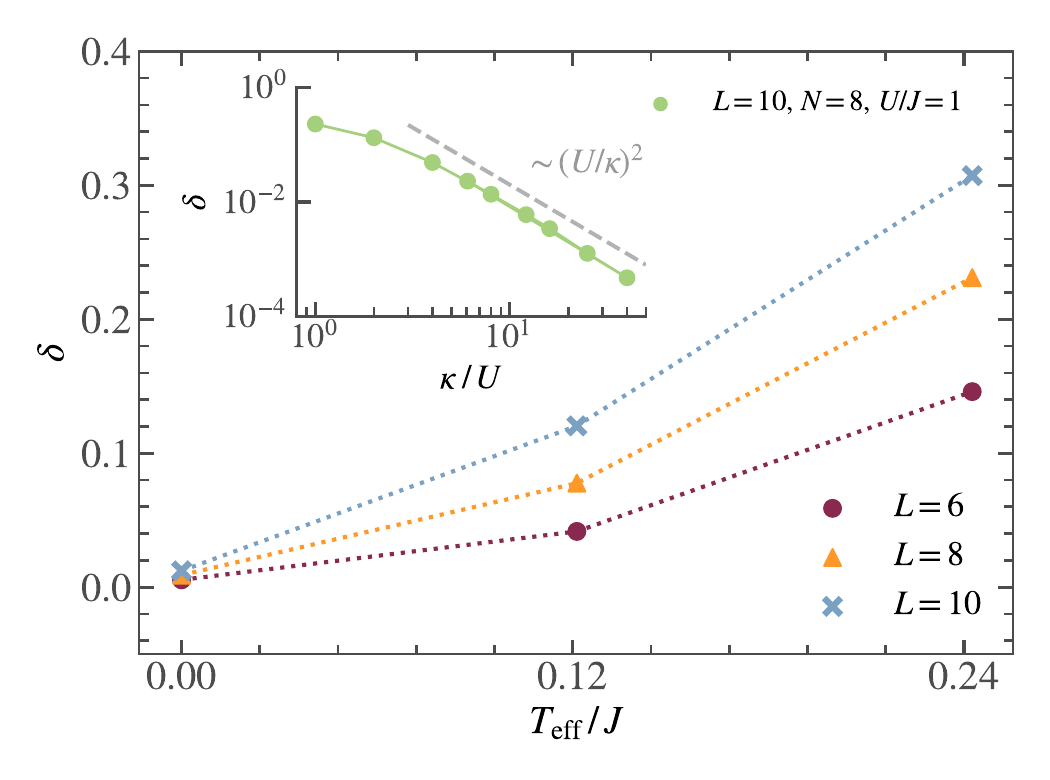}
    }
    \vspace{-0.2cm}
    \caption{
    Simulating the dissipative preparation of a \gls{BEC}. We show the off\hyp diagonal element $\bm \gamma_{1L}$ of the one\hyp body density matrix for several $\kappa/U$. We recover the scaling \cref{eq:app:long:range:order}, as predicted by Bogoliubov theory. This shows that the steady state features a lattice analogue of long\hyp range order.
    Inset: Off\hyp diagonal elements of the one\hyp body density matrix for two different $\kappa/U$.
    All calculations were performed with the parameters $L=10$, $N=8$, $U=1$ and $J=1$, as well as $k_0 = 0$. The local dimension was set to $d=N+1$.
    \updated{Panel b):  The condensate depletion as a function of the effective temperature defined in \cref{eq:app:effective:temp}.
    All datasets were obtained at unit filling $N/L=1$.
    Inset: $\kappa$\hyp dependence of the depletion. We recover the $\kappa^{-2}$ scaling predicted by Bogoliubov theory.
    In both insets, the couplings are $U=J=1$ with local dimension $d=N+1$.
    In the upper inset, $L=8$ and $N=10$, while in the second one $L=10$ and $N=8$ were chosen.
    Except for the second inset, we always set $\kappa=2J$.}
    }
    \label{fig:app:long:range:order}
\end{figure}
\updated{
Then, in \cref{subfig:depletion:T:dependence}, we show that the condensate depletion increases with the effective temperature $T_{\text{eff}}$.
In fact, if we take a typical value $J=1\mathrm{kHz}$ for ultracold atoms,~\cref{subfig:depletion:T:dependence} indicates the formation of a condensate with depletion $\delta\sim 0.1$ for temperatures of about $10 \mathrm{nK}$.
Note that the inset confirms the validity of our perturbative arguments: Even for moderate values of the dissipation strength, we observe the expected scaling of the depletion $\sim (U/\kappa)^2$.}
\section{Numerical implementation of the dissipative dynamics} \label{sec:app:numerical:details}

Here, we provide the details of the numerical implementation of the dissipative time evolution considered in the main text.
We employ the \gls{MPS} representation \cite{Schollwoeck2011} and vectorize the density matrices by doubling the system size, alternating physical and auxiliary sites, as shown schematically in \cref{fig:app:vectorized:sketch}. 
This arrangement of the sites avoids introducing long\hyp range terms \cite{Casagrande2021} in the \gls{MPO} representation of the vectorized Lindbladian \cite{Wolff2020} (see \cref{eq:app:Lindblad:vectorized}).
We compute the dissipative dynamics employing the \gls{TDVP} method for \gls{MPS} \cite{Haegeman2011,Paeckel2019}.
For highly\hyp excited bosonic systems, characterized by large local physical dimension $d$, the most effective variant is the \gls{LSE-TDVP} \cite{Yang2020,Grundner2023}. 
This is based on single\hyp site updates combined with a local subspace expansion \cite{Hubig2015}, and is thus faster than the conventional \gls{2TDVP} by a factor of $d$.
We emphasize that care needs to be taken when using \gls{TDVP} for Lindbladians, since they are non\hyp Hermitian. 
In this context, two possible strategies are to either decompose the Lindbladian into a Hermitian and an anti\hyp Hermitian part and to alternate real and imaginary Trotterized time steps or to perform a brute\hyp force Taylor expansion of the exponentials of the local site tensors, as detailed in \cite{Moroder_diss}. We followed the latter strategy.

To compute the dissipative dynamics efficiently, it is imperative to exploit the system's symmetries.
First, note that the Lindbladian specified by the unitary part \cref{eq:app:bose:hubbard:ham} and the dissipative component \cref{eq:app:jump:ops} conserve the total number of particles, and thus possess a corresponding global $U(1)$ symmetry. 
Moreover, since neither coherent nor dissipative hopping exchange particles between the physical and auxiliary sites (see panel a in \cref{fig:app:vectorized:sketch}), $\mathcal{L}$ possesses a second $U(1)$ symmetry associated with the particle conservation on the sublattices. This second symmetry only exists in the vectorized system and guarantees, that physical states are not connected to unphysical ones, which appear due to the enlargement of the Hilbert space.
We exploit both the symmetries to obtain a block\hyp decomposition of the Lindbladian.
%
%Simulating the dynamics of an N\hyp boson state is simplified greatly through the present $U(1)$ symmetries.
%
Furthermore, these symmetries ensure that at most $N$ particles can sit on each site in the vectorized lattice (though $2N$ particles are on the lattice in total), allowing us to choose local dimension $d = N+1$.
\begin{figure}
    \centering
    \subfloat[\label{fig:app:vectorized:a}]{
    \includegraphics[width=0.95\columnwidth]{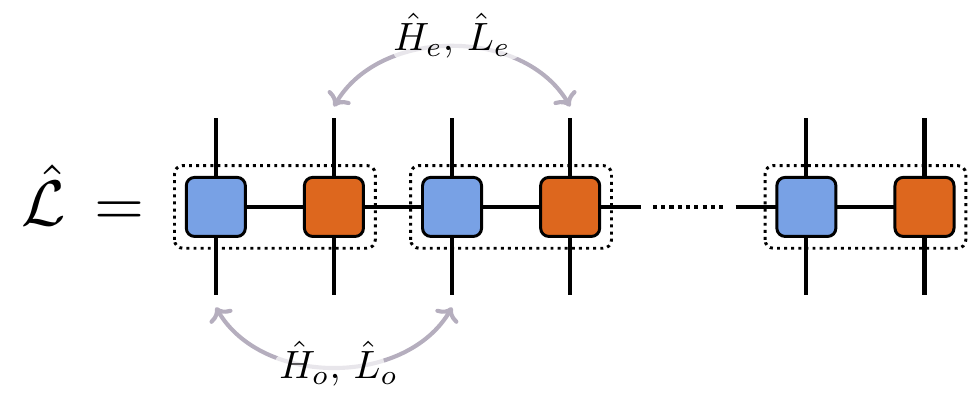}
 }
    \vspace{-0.2cm} % Add some vertical space between subfigures
    % Panel (c)
    \subfloat[\label{fig:app:vectorized:b}]{
    \includegraphics[width=0.95\columnwidth]{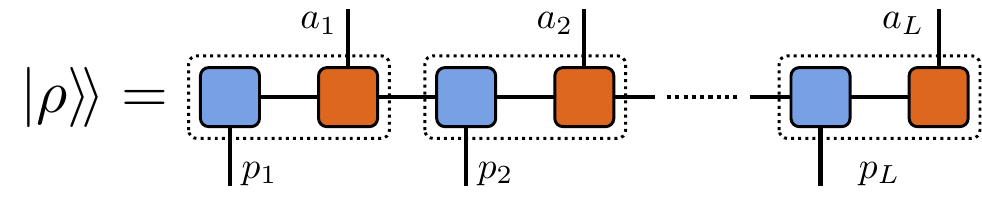}
 }
\caption{The \gls{MPO} representation of a vectorized Lindbladian (panel a) and of the \gls{MPS} representation of a vectorized density matrix (panel b). Vectorization is carried out by purifying the system, i.e. adding an auxiliary site, marked in orange, for every physical site (colored blue). To ensure that the Lindbladian has only local terms, physical and auxiliary sites are alternated. Terms in \cref{eq:app:Lindblad:vectorized} that are on the left (right) of a tensor product act on odd  (even) sites respectively and are indexed by o (e) accordingly. For the dissipative dynamics of a bosonic system that we consider here, the Hamiltonian and the jump operators feature next\hyp nearest neighbor hopping.}
    \label{fig:app:vectorized:sketch}
\end{figure}

Due to the strong increase in bond dimension in the first few time steps, especially when starting from a pure product state, we initially use a small time step of $5\times10^{-4}$ and increase it after 20 steps to either $0.01$ or $0.005$ (for large scale calculations also 0.002), depending on system size.
We track the norm of the time\hyp evolved state, which needs to stay smaller than 1 to ensure a proper, physical evolution.
We also track the maximal truncation error, which is defined as the maximal discarded weight \cite{Paeckel2019} on a single site.
In \cref{fig:app:trunc:and:bond} we display the bond dimension (right axis) and the maximal truncation error (left axis) for one dissipative evolution considered in the main text.
Notice that due to the symmetries discussed in \cref{sec:app:unitary:equivalence}, it is enough to perform calculations at zero characteristic momentum $k_0$.
All calculations were performed using the \textsc{SyTen} toolkit \cite{hubig:_syten_toolk,hubig17:_symmet_protec_tensor_networ}.
\begin{figure}
    \centering
    \includegraphics[width=\columnwidth]{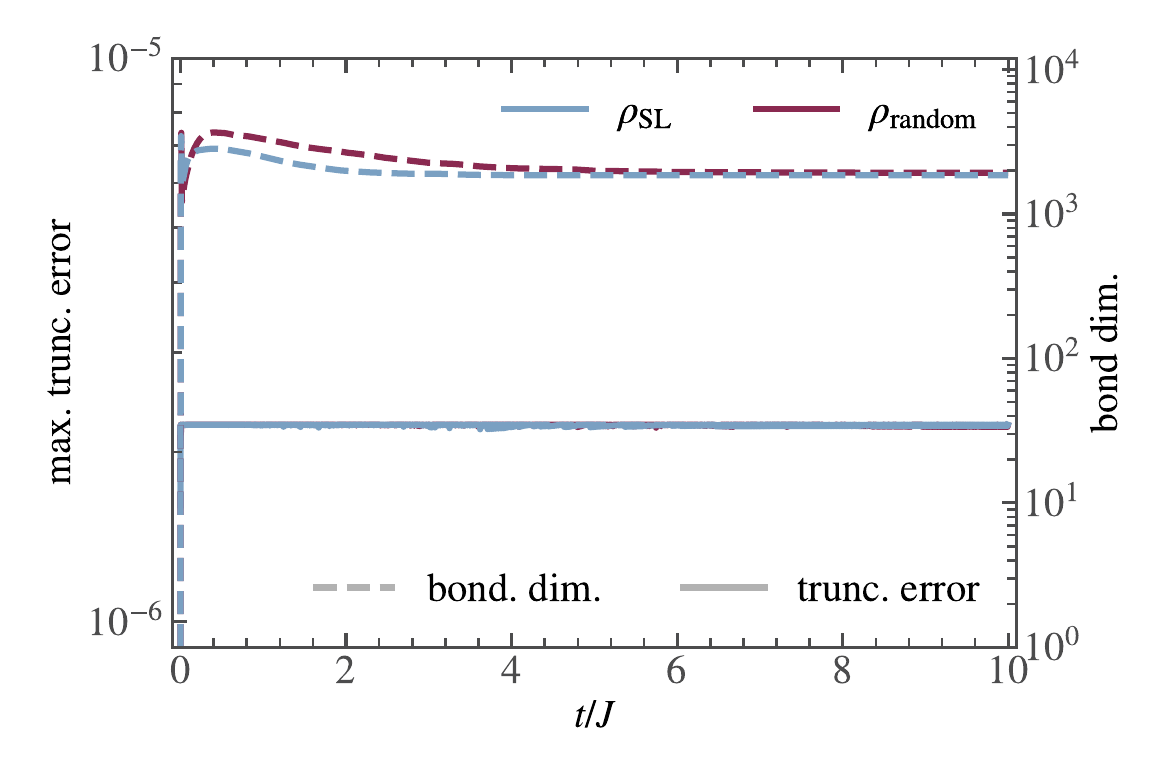}
    \caption{Truncation error and right bond dimension for the Bose\hyp Hubbard model with 10 sites and 10 particles, with $U/J=0$ and $\kappa/J=2$. For both initial states, there is a massive increase in bond dimension in the first time steps; later, it saturates at about 2000. To accurately capture the changes in the first steps, the time step was initially set to $5\times10^{-4}$, and after 20 steps it was increased to 0.005. The truncation error stays constant at about $2\times 10^{-6}$.}
    \label{fig:app:trunc:and:bond}
\end{figure}

Many studies of the Mpemba effect based on \gls{ED}, employ the quantum relative entropy as a (pseudo) measure of distance from the steady state, whose exponential decay follows the exponent of the slowest\hyp decaying mode \cite{Moroder2024,Graf2025}.
However, such a quantity is very difficult to compute in the \gls{MPS}\hyp vectorized framework. 
The only distance measure that is straightforward to compute in this picture is the $L_2$\hyp norm
\begin{equation}
 \lVert\hat{\rho}_1-\hat{\rho}_2\rVert_2^2 = \text{Tr}\big{(} (\hat{\rho}_1-\hat{\rho}_2)^2 \big{)} = \vecbraket{\hat{\rho}_1-\hat{\rho}_2}{\hat{\rho}_1-\hat{\rho}_2} \;.
\end{equation}
This norm will show the exponential equilibration accurately.
To see this, assume $\lambda_3<\lambda_2$ and note that at a late time $t \gg 1/|\text{Re}(\lambda_2)|$, the time evolved state is accurately approximated by  $\hat{\rho}(t) =\hat{\rho}_{\text{ss}}+a_2\mathrm e^{\lambda_2 t}\hat{r}^\nodagger_2+ \mathcal{O}\big(\mathrm e^{t \, \text{Re}(\lambda_3)}\big)$. 
Thus, we immediately get that
\begin{equation}
 \lVert\hat{\rho}(t)-\hat{\rho}_{\text{ss}}\rVert_2 = \mathrm e^{t \, \text{Re}(\lambda_2)} |a_2| \; \lVert \hat{r}^2_2 \rVert_2 + \mathcal{O}\big(\mathrm e^{t \, \text{Re}(\lambda_3)}\big)\; ,
 \label{eq:app:distance:from:ss}
\end{equation}
which is exponential with rate $\text{Re}(\lambda_2)$.
This makes it possible to fit the exponential convergence of a slowly decaying initial state, in order to predict the long time behavior of such states, circumventing the necessity to perform the whole time evolution. 
Note that this procedure (i.e. linearly fitting the logarithm of \cref{eq:app:distance:from:ss}) was used to find the relaxation times of some of the random states  used for the calculation of the speedups in Fig. 3b) in the main text.

For all our \gls{MPS}\hyp calculations, we obtain the steady state by a long\hyp time evolution.
Although this leads to an approximate steady state $\hat{\rho}_{\text{ss}} \approx\hat{\rho}(t_{\text{max}})$, we can track if it is sufficiently converged, by seeing if $\lVert\hat{\rho}(t)-\hat{\rho}_{\text{ss}}\rVert_2$ stays the same when considering $\hat{\rho}(t_{\text{max}})$ and $\hat{\rho}(t_{\text{max}}-\Delta t)$ as the steady state for a sufficiently big $\Delta t$. A similar analysis can be performed by comparing the leading eigenvalue $N_0$ of the one\hyp body density matrix $\bm \gamma$ for $t_{\text{max}}$ and $t_{\text{max}}- \Delta t$.

The enforcing of normalization and computation of expectation values for density matrices presents another difficulty: In the vectorized picture, the normalization corresponds to
\begin{equation}
    1 = \text{Tr}(\hat{\rho}) = \vecbraket{\hat{\mathds{1}}}{\hat{\rho}} \, .
\end{equation}
Unfortunately, $\vecket{\mathds{1}}$ is a strongly entangled state and thus requires high bond dimensions.
\begin{figure}
    \centering
        \includegraphics[width=\columnwidth]{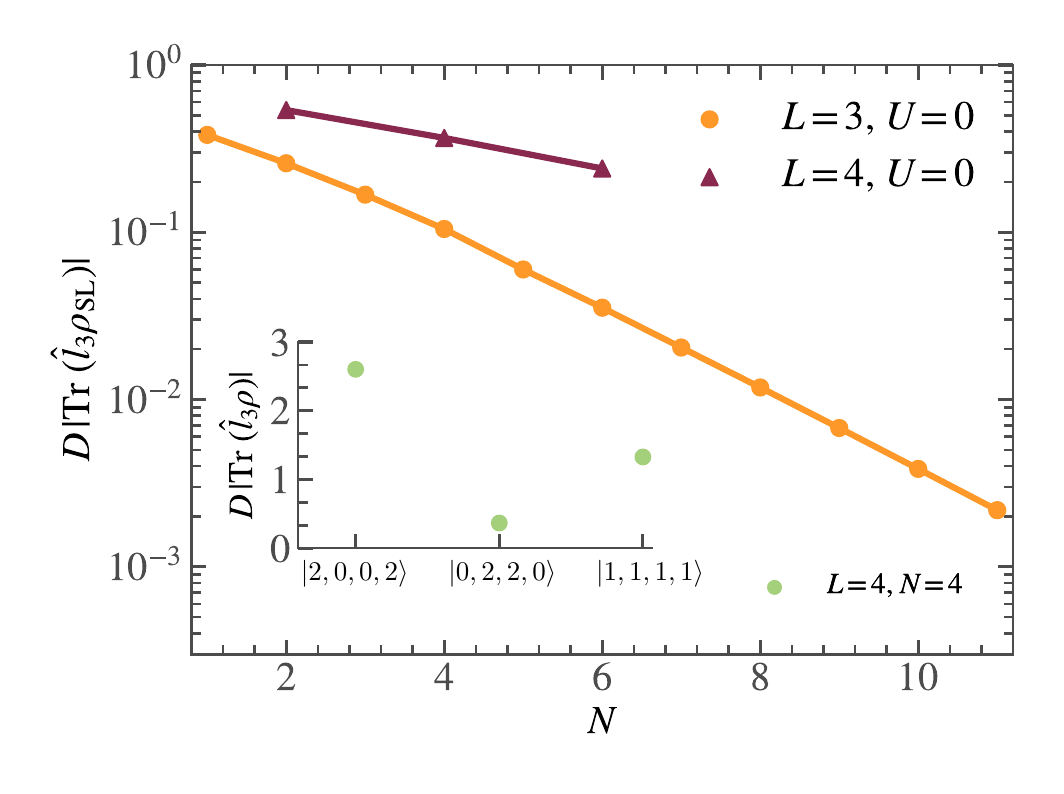}
    \caption{Normalized overlap of the \acrfull{SL} state with  the second\hyp slowest\hyp decaying mode $\hat l_3$.
    The overlap vanishes exponentially with the number of bosons $N$, and thus, the convergence rate gets better upon increasing $N$.
    Inset: Overlap of the different symmetric states with $\hat l_3$ for the system specification $L=4$, $N=4$.
    The \gls{SL}\hyp state (i.e. $\ket{0, 2, 2, 0}$) has the smallest overlap with $\hat l_3$, making it the ideal initial state among all symmetric states. Note that also the homogeneous state $\ket{1, 1, 1, 1}$ has low overlap.
    For all calculations, we used the parameters $\kappa=J$ and $U=0$ and \gls{ED} to find the eigenmodes.}
    \label{fig:app:overlap:l3}
\end{figure}
However, we can make use of the $U(1)$\hyp symmetries and only calculate the state in one particular particle sector \cite{Paeckel2019}.
We define the operator
\begin{equation}
    \hat{C}_{\text{tot}} = \sum_{j=1}^{L} \hat{b}_{2j}^\dagger \hat{b}_{2j+1}^\dagger \;.
\end{equation}
Then the vectorized identity for $N$ particles is given by 
\begin{align}
    \vecket{\mathds{1}} = \frac{1}{N!}\hat{C}_{\text{tot}}^N\ket{\mathrm{vac}} \;.
\end{align}
This relation is easily proven using the multinomial theorem.

Finally, we focus on the equilibration behaviour of different symmetric initial states.
As discussed in \cref{sec:app:fast:states}, we identified a class of exponentially faster\hyp equilibrating states, namely those that are invariant under reflections about the center of the lattice. 
We are mainly interested in those that are also pure\hyp product states, as they are particularly useful in experimental realizations.
Among these states, the fastest\hyp equilibrating one is the symmetric state that has the lowest overlap with the second\hyp slowest decaying mode $\hat l_3^\nodagger$.
To be able to compare different system specifications with each other quantitatively, we need to normalize $\hat l_3$, such that the sum over absolute values of the diagonal elements is equal to one. 
This is done by multiplying with the Hilbert space dimension $D$, which leads to a comparable quantity that fulfills
\begin{equation}
 D\big{|}\text{Tr}(\hat l_3\hat{\rho}_{\text{random}})\big{|} = 1 \, ,
\end{equation}
when averaging over many $\hat{\rho}_{\text{random}}$.
In \cref{fig:app:overlap:l3}, with the help of \gls{ED}, we show that the overlap of the \gls{SL} state (having all bosons located on the central sites) with $\hat{l}_3$ decays exponentially with $N$.
This helps explaining why in Fig. 3b in the main text the speedups remain approximately constant upon increasing the system size, despite the fact that the Lindbladian spectrum becomes denser.
Moreover, the inset indicates that the \gls{SL} state has the smallest overlap with $\hat{l}_3$ among all symmetric states.
\updated{In the main text we introduced the notion of speedup $S(\epsilon)$.
It was used to quantify the relative reduction of time to get to prepare the steady state up to a distance $\epsilon$, between a symmetric and random intial state.
Rigorously speaking, we can define these speedups $S(\epsilon)$ of a \gls{SL} state is given by }
\begin{equation}
    \updated{S(\epsilon) = \frac{t(\epsilon, \hat\rho_\mathrm{random})}{t(\epsilon, \hat\rho_\mathrm{SL})}\; , \hspace{0.4cm} t(\epsilon, \hat\rho) = \min\big\{ t\, \big\vert \, \Vert \hat\rho(t)-\hat\rho_\mathrm{ss}\Vert_2 \leq \epsilon\big\}\, .}
    \label{eq:bec:speedup:ratio:experiment}
\end{equation}
\updated{We are most interested in the small $\epsilon$ behavior of $S(\epsilon)$. If $\epsilon\ll 1$ the stopping times satisfy $t(\epsilon, \hat\rho)\gg -1/\mathrm{Re}(\lambda_2)$ and we have $\Vert \hat\rho(t)-\hat\rho_\mathrm{ss}\Vert_2 = \vert\vecbraket{l_j}{\rho}\vert \mathrm e^{t\mathrm{Re} \lambda_j}$, where $j=2$ for a random and $j=3$ for the \gls{SL} state. Rearranging for $t(\epsilon, \hat\rho)$ and inserting into \cref{eq:bec:speedup:ratio:experiment} yields}
\begin{equation}
    \updated{S(\epsilon) \xrightarrow[]{\epsilon\ll1} \, \frac{\mathrm{Re} \, \lambda_3}{\mathrm{Re} \, \lambda_2} \; ,}
    \label{eq:app:limit:of:speedup}
\end{equation}
\updated{which shows the connection to the spectral properties of the Lindbladian. Notice, that if the overlap of the slowest decaying mode with $\hat l_3$ is very small (c.f. the exponential decrease in \cref{fig:app:overlap:l3}), $S(\epsilon)$ will first approach $\mathrm{Re} \, \lambda_4/\mathrm{Re} \, \lambda_2$ at intermediate $\epsilon$, after finally showing the limit behavior \cref{eq:app:limit:of:speedup}.}
\bibliography{Literature}